\documentclass[a4paper,10pt,twocolumn]{article}

\usepackage{amsmath,amsfonts,mathtools}
\usepackage{amssymb}

\usepackage[T1]{fontenc}
\usepackage{lmodern}
\usepackage{bm}
\usepackage{etoolbox}
\AtBeginEnvironment{equation}{\small}
\AtBeginEnvironment{equation*}{\small}
\AtBeginEnvironment{align}{\small}
\AtBeginEnvironment{align*}{\small}
\AtBeginEnvironment{multline}{\small}
\AtBeginEnvironment{multline*}{\small}
\allowdisplaybreaks

\usepackage[a4paper,margin=1.8cm,columnsep=0.7cm]{geometry}
\usepackage{setspace}
\usepackage{graphicx}
\usepackage{booktabs}
\usepackage{multirow}
\usepackage{array}
\usepackage{caption}
\usepackage{stfloats}
\usepackage{cuted}
\usepackage{placeins}

\usepackage{listings}
\usepackage{xcolor}
\usepackage{textcomp}
\usepackage[numbers,sort&compress]{natbib}
\usepackage{hyperref}
\hypersetup{
  unicode=true,
  colorlinks=true,
  linkcolor=blue,
  citecolor=blue,
  urlcolor=blue,
  pdftitle={Maximum Entropy Encoding of Energy-Weighted Spherical Moments},
  pdfauthor={Jiaze Sun},
  pdfkeywords={maximum entropy, spherical moments, irradiance reconstruction, spherical harmonics, real-time rendering, compact lighting representation}
}

\DeclareMathOperator{\Tr}{Tr}
\DeclareMathOperator{\TF}{TF}
\DeclareMathOperator{\Cov}{Cov}
\DeclareMathOperator{\Var}{Var}
\DeclareMathOperator{\diag}{diag}

\newcommand{\RR}{\mathbb{R}}
\newcommand{\EE}{\mathbb{E}}

\newcommand{\norm}[1]{\lVert#1\rVert}
\newcommand{\calN}{\mathcal{N}}

\newcommand{\calC}{\mathcal{C}}
\newcommand{\dif}{\mathrm{d}}
\newcommand{\etal}{\textit{et al.}}

\title{\vspace{-1.5cm}\bfseries\LARGE Maximum Entropy Encoding of Energy-Weighted Spherical Moments}
\author{Jiaze Sun\\
{\small Northwestern Polytechnical University}\\[2pt]
{\small \texttt{sjz1@mail.nwpu.edu.cn}}}
\date{}

\begin{document}

\maketitle

\begin{abstract}
\noindent
We study how angular energy signals composed of non-negative Monte Carlo path samples can be compressed and reconstructed for irradiance using finite moments. Writing each sample as an energy-weighted directional feature $\mathbf{x}=r\mathbf{u}$, we adopt total energy, the first directional moment, and the traceless second moment as $1+3+5$ linearly additive, rotationally covariant statistics. Under a fixed Lebesgue reference measure, the maximum-entropy closure yields $p(r,\mathbf{u})\propto\exp(-\beta r g(\mathbf{u})),\;g(\mathbf{u})=1-\mathbf{b}\cdot\mathbf{u}+\mathbf{u}^{\mathsf{T}}\mathbf{Q}\mathbf{u}$, whose directional probability and angular energy density are proportional to $g^{-3}$ and $g^{-4}$, respectively. When $g_{\min}>0$ the closure is normalizable and the reconstruction is strictly positive. We further provide analytic moment matching, variance, inverse sampling, and closed-form diffuse response for the pure-dipole four-parameter subfamily, as well as the realizability domain, partition function, azimuthal algebraic integral, and LUT-oriented reconstruction form for the dipole--second-moment coaxial five-parameter subfamily.

Experiments cover 981 Poly Haven HDRI~2K scenes and three Debevec probes. Five-parameter MaxEnt achieves a 78.7\% per-scene win rate against stored QZH (bootstrap 95\% CI excludes zero), with mean luminance RMSE reduced by 15.8\%; the advantage is more pronounced in scenes with strong directionality and high second-band energy fraction (91.3\% win rate across the 392 scenes with $\rho\geq 0.6$). Both MaxEnt variants maintain zero negative irradiance across all scenes, while SH and QZH exhibit negative values on 52--579 scenes. On the natural HDR benchmark, full second-order SH-2 (9~DOF) yields the lowest overall error, with five-parameter MaxEnt ranking second and outperforming SH-2 in the high-directionality bucket; the coaxial subfamily shows systematic closure error on non-coaxial multi-source scenes.
\end{abstract}

\noindent\textbf{Keywords:} maximum entropy, spherical moments, irradiance reconstruction, spherical harmonics, real-time rendering, compact lighting representation


\section{Problem Definition, Contributions, and Notation}

\subsection{Problem Definition}

The input is a set of non-negative weighted directional samples $\mathbf{x}_i=L_i\mathbf{u}_i$ per color channel, where $L_i\geq 0$ and $\mathbf{u}_i\in S^2$. The goal is to store angular energy moments up to $l\leq 2$ in a finite-dimensional state that supports linear accumulation and spatial rotation, and to reconstruct a non-negative directional energy density and normalized cosine irradiance from these moments. Signed estimators and negative filter weights require a positive--negative measure decomposition, which lies outside the scope of the single non-negative measure model adopted here.

\subsection{Main Contributions}

\begin{itemize}
  \item We give a set of nine scalar encodings linearly equivalent to the full real $l\leq 2$ angular energy moments, and identify the representation class for which this dimensional lower bound is attained;
  \item From the independent $1+3+5$ moment constraints we derive the MaxEnt closure with $g^{-3}$ directional probability and $g^{-4}$ energy density, giving normalizability conditions and the relationship to SH projection;
  \item On the existing Levermore M1 pure-dipole closure we derive graphics-oriented analytic moment matching and cosine response, and for the coaxial five-parameter subfamily we provide moment-space projection, partition function, and one-dimensional diffuse reconstruction;
  \item We establish the low-frequency correspondence between five-parameter moments and stored QZH, and compare accuracy and negative-value behavior on a 981-scene HDR benchmark and analytic stress tests.
\end{itemize}

\subsection{Notation and Irradiance Conventions}

We denote the angular energy density by $\ell(\mathbf{u})$ and use
\begin{equation*}
  E_{\text{raw}}(\mathbf{n})=\int_{S^2}\ell(\mathbf{u})\max(\mathbf{n}\cdot\mathbf{u},0)\,\dif\Omega,
  \qquad
  E(\mathbf{n})=\frac{E_{\text{raw}}(\mathbf{n})}{\pi}.
\end{equation*}

Experiments and shaders labeled ``normalized cosine'' adopt $E$; when discussing analytic hemispherical integrals we use $E_{\text{raw}}$ explicitly. Throughout the paper we refer to $l\leq 1$ and $l\leq 2$ as SH-1 and SH-2, respectively, by their maximum spherical harmonic degree; the ZH3 paper~\cite{Roughton2024ZH3} refers to them as SH2 and SH3 by band count.

\section{Related Work}

Low-order spherical harmonics are a classic representation for diffuse environmental lighting. Ramamoorthi and Hanrahan provided the low-frequency SH analysis of irradiance, and Sloan summarized common practical techniques for real-time implementations~\cite{Ramamoorthi2001Efficient,Sloan2008StupidSH}. Roughton~\etal's ZH3 augments linear SH with a single axisymmetric second-order coefficient and studies storage, inference, and shared-axis strategies~\cite{Roughton2024ZH3}. The present work shares the same low-order moment coordinates but employs a maximum-entropy closure to produce a strictly positive nonlinear density.

The maximum entropy principle was systematized by Jaynes as the principle for choosing a distribution under given constraints and a reference measure~\cite{Jaynes1957MaxEnt,Jaynes1957MaxEntII}. In radiative transfer, Minerbo constructed Eddington factors from maximum entropy, and Levermore gave the closure commonly known as M1 for energy--flux moments~\cite{Minerbo1978,Levermore1984}; the latter can also be interpreted as the energy--momentum tensor of massless radiation that is isotropic in a particular rest frame after a Lorentz boost. The pure-dipole angular energy family $\ell(\mathbf{u})\propto(1-\mathbf{b}\cdot\mathbf{u})^{-4}$ coincides with the angular distribution of Levermore's M1, so we do not claim it as a new transport closure; our contributions for this subfamily are the graphics-oriented moment-encoding interpretation, analytic cosine response, and runtime formulas.

Spherical second moments are often represented by the Bingham or the more general Fisher--Bingham exponential family~\cite{Bingham1974,Kent1982}. The difference here is that samples additionally carry a non-negative radial weight, and the constraint uses $\EE[R\mathbf{U}\mathbf{U}^{\mathsf{T}}]$ rather than the unweighted second moment of the spherical probability. This radial coupling yields directional probability $\propto g^{-3}$ and energy density $\propto g^{-4}$. The five-parameter coaxial model that preserves second angular moments is structurally related to second-order maximum-entropy moment closures (M2) in radiative transfer~\cite{Sarr2020M2}, but our entropy functional, reference measure, and coaxial restriction are not identical term-by-term to standard M2; we therefore claim only a structural connection.

Spherical Gaussians (SG) and their anisotropic extensions are widely used for compact lighting and integration approximations~\cite{Xu2013ASG,Tokuyoshi2022SG}. A single-lobe SG couples its first and second moments through a single concentration parameter, whereas our five-parameter subfamily independently stores one axial second-order moment. Silvennoinen~\etal~\cite{Silvennoinen2025SHE} recently introduced Spherical Harmonic Exponentials for efficient glossy reflections (HPG~2025), which fit a factorized spherical-harmonic exponential representation in log space for glossy-reflection reconstruction; our work is complementary in that we derive a moment-preserving MaxEnt closure from first principles rather than fitting an exponential to a target radiance distribution. Path guiding and spatiotemporal denoising are potential applications, but the present experiments focus on representation and irradiance reconstruction; full task-level evaluation requires separate measurements of sampling variance, temporal stability, and runtime cost.


\section{First Principles}

\textit{Note: This paper defines Monte Carlo path samples as energy-weighted directional estimators with non-negative weights. The BRDF is evaluated later; ``energy'' below refers to the scalar weight used for reconstruction, and the four-dimensional feature structure serves the unified encoding of these weights and their directional moments.}

In a path-tracing pipeline with deferred BRDF evaluation, a single sample can be written as the product of a non-negative scalar weight and a direction. Given a set of energy-weighted directional samples $\{\mathbf{x}_i\}_{i=1}^n$ where $\mathbf{x}_i=L_i\hat{\mathbf{x}}_i\in\RR^3$, $L_i\geq 0$ is a scalar that incorporates the sampling weight convention, and $\hat{\mathbf{x}}_i$ is a unit direction, our goal is to find a compressed representation that encodes the distribution of these weighted directional samples. If an implementation uses signed estimators or negative filter weights, the positive and negative measures must first be decomposed; the cone constraints below cannot be applied directly.

We adopt the following testable encoding design constraints:
\begin{itemize}
  \item Non-negative weights: $L_i\geq 0$;
  \item First-order homogeneity in energy scale: uniformly scaling all samples scales the encoding identically;
  \item Linear additivity: the encoding of a collection of samples can be accumulated term-by-term;
  \item Rotational covariance: coordinate rotations transform the encoding according to the corresponding representation.
\end{itemize}

Under these conventions, the state of a single sample is described by $(|\mathbf{x}_i|,\mathbf{x}_i)$. The temporal component of the four-dimensional feature records the scalar weight and the spatial components record the directionally-weighted quantity; the moment structure of this feature is studied below in an $\operatorname{SO}(3)$ rotationally covariant framework.

These constraints provide a unified structural foundation for filtering, moment matching, and irradiance reconstruction. In addition, we require the compressed representation to:
\begin{itemize}
  \item Preserve angular moments up to second order completely, so as to express anisotropic shapes;
  \item Be first-order homogeneous in sample weight, linearly additive over sample collections, and covariant under rotation;
  \item Use the fewest degrees of freedom among representation classes satisfying these constraints;
  \item Under given moment constraints and a reference measure, select the distribution that introduces no unobserved information via the maximum entropy principle.
\end{itemize}

\textit{Note: ``Minimum'' and ``maximum entropy'' here each have a precise scope: the minimum number of degrees of freedom is nine real parameters, and nine is the dimensional lower bound for any linear, rotationally covariant representation that fully preserves spherical $l\leq 2$ moments; maximum entropy is conditional optimality under the specified constraints and the Lebesgue reference measure. Neither claim implies that no better scheme exists under a different encoding, a different loss function, or a different reference measure.}


\section{Derivation of a Four-Dimensional Tensor Representation}

Under the framework of first-order homogeneity, linear additivity, rotational covariance, and complete second-order angular moment preservation, expanding each sample into a four-dimensional energy--direction feature and taking its normalized outer product yields a natural minimum-dimensional representation. The four-dimensional object employs Euclidean feature packing to jointly record scalar energy, the first directional moment, and the second angular moment.

Write the weighted directional output of a single $1\,\text{spp}$ ray as a four-dimensional energy--direction feature:
\begin{equation*}
  k_i^\mu = \begin{pmatrix}k_i^0 \\ \mathbf{k}_i\end{pmatrix}
  = \begin{pmatrix}|\mathbf{x}_i| \\ \mathbf{x}_i\end{pmatrix} \in \RR^4.
\end{equation*}

To extract second-order information about the angular distribution algebraically, consider the outer product of $k_i^\mu$ with itself, reduced by $k_i^0$ to preserve first-order homogeneity:
\begin{equation*}
  \tau_i^{\mu\nu} = \frac{k_i^\mu k_i^\nu}{k_i^0}
  = \frac{k_i^\mu k_i^\nu}{|\mathbf{x}_i|}.
\end{equation*}

This expression holds directly when $\mathbf{x}_i\neq\mathbf{0}$; for zero-energy samples we adopt the continuous extension $\tau_i^{\mu\nu}=0$. This avoids $0/0$ while preserving first-order homogeneity and linear additivity. Expanding $\tau_i^{\mu\nu}$ component-wise:

\begin{itemize}
  \item $00$ component: $\tau_i^{00}=|\mathbf{x}_i|=L_i$, the total energy scalar $\omega$.
  \item $0j$ components ($j=1,2,3$): $\tau_i^{0j}=x_i^j$, the direction vector $\mathbf{x}_i$, which is precisely the dipole momentum $\mathbf{v}$.
  \item $jk$ components ($j,k=1,2,3$): $\tau_i^{jk}=L_i(\hat{\mathbf{x}}_i\otimes\hat{\mathbf{x}}_i)$, the normalized directional outer product matrix.
\end{itemize}

Thus $\tau_i^{\mu\nu}$ is a $4\times 4$ real symmetric tensor with ten apparent independent components. However, the trace of the spatial block is identically equal to the $00$ component:
\begin{equation*}
  \Tr(\tau_i^{jk}) = \sum_{j=1}^3 \frac{x_i^j x_i^j}{|\mathbf{x}_i|}
  = |\mathbf{x}_i| = \tau_i^{00}.
\end{equation*}

This algebraic identity follows from the unit-direction constraint $|\hat{\mathbf{x}}_i|=1$, so $\tau_i^{\mu\nu}$ has only $10-1=9$ independent degrees of freedom.

The total aggregate state is obtained by algebraic accumulation over all samples: $T_{\text{total}}^{\mu\nu}=\sum_{i=1}^n \tau_i^{\mu\nu}$. $T_{\text{total}}^{\mu\nu}$ is the desired additive compressed representation, with the following degree-of-freedom decomposition:
\begin{itemize}
  \item $\tau^{00}$: total energy $\omega$ (1 DOF)
  \item $\tau^{0j}$: dipole momentum $\mathbf{v}$ (3 DOF)
  \item Traceless part $\mathbf{D}=\TF(\tau^{jk})$: anisotropic shear stress moment (5 DOF)
\end{itemize}

Total: $1+3+5=9$ DOF. Here and below, $\mathbf{D}$ denotes the stress in moment space; $\mathbf{Q}$ (introduced later) denotes the corresponding MaxEnt natural parameter linked to $\mathbf{D}$ through nonlinear moment matching---the two are not component-wise equal. This tensor simultaneously satisfies physical conservation, second-order completeness, a number of DOF attaining the dimensional lower bound (the dimension of the spherical $l\leq 2$ moment space is exactly $1+3+5=9$), and linear additivity.

\subsection{Necessity of the Moment Constraints and Mismatch of Alternative Closures}

This subsection rigorously answers why the second-order constraint must employ the energy--stress moment defined above, and cannot be replaced by the classical statistical second moment with its Gaussian MaxEnt closure, or by a single-lobe spherical Gaussian.

\subsubsection{Target Object and Definition of ``Strict Degeneracy''}

Let the random weighted directional sample be $\mathbf{X}=R\mathbf{U}$, with $R=|\mathbf{X}|\geq 0$ and $\mathbf{U}\in S^2$. The object used for radiance, spherical harmonic projection, and deferred BRDF reconstruction is the angular energy density:
\begin{equation*}
  \ell(\mathbf{u}) = \int_0^\infty r\,p(r\mathbf{u})\,r^2\dif r
  = \int_0^\infty r^3 p(r\mathbf{u})\dif r.
\end{equation*}

The corresponding first three angular moments are
\begin{align*}
  \omega &= \int_{S^2}\ell(\mathbf{u})\dif\Omega = \EE[R],\\
  \mathbf{v} &= \int_{S^2}\mathbf{u}\,\ell(\mathbf{u})\dif\Omega = \EE[R\mathbf{U}] = \EE[\mathbf{X}],\\
  \mathbf{S} &= \int_{S^2}\mathbf{u}\mathbf{u}^{\mathsf{T}}\ell(\mathbf{u})\dif\Omega
  = \EE[R\mathbf{U}\mathbf{U}^{\mathsf{T}}]
  = \EE\!\left[\frac{\mathbf{X}\mathbf{X}^{\mathsf{T}}}{|\mathbf{X}|}\right].
\end{align*}

Since $|\mathbf{U}|=1$, $\Tr(\mathbf{S})=\omega$. Hence the independent second-order angular information is exactly $\mathbf{D}=\TF(\mathbf{S})=\mathbf{S}-\frac{\omega}{3}\mathbf{I}$ with $\Tr(\mathbf{D})=0$.

By ``exact reduction to $l\leq L$ spherical harmonics'' we mean that the reconstructed finite band matches the orthogonal projection $\Pi_L[\ell]$ exactly, i.e., all $l\leq L$ coefficients are preserved. To avoid naming ambiguity, we denote $l\leq 1$ as SH-1 and $l\leq 2$ as SH-2 by maximum degree.

\subsubsection{General Second-Order Orthogonal Projection Theorem}

Any real $l\leq 2$ function can be uniquely written as
\begin{equation*}
  q(\mathbf{u}) = a + \mathbf{c}\cdot\mathbf{u} + \mathbf{u}^{\mathsf{T}}\mathbf{C}\mathbf{u},
  \quad \mathbf{C}=\mathbf{C}^{\mathsf{T}},\; \Tr(\mathbf{C})=0.
\end{equation*}

Using spherical symmetry integrals and noting that odd-degree monomial integrals vanish, we obtain
\begin{align*}
  \int_{S^2} q\dif\Omega &= 4\pi a,\qquad
  \int_{S^2} \mathbf{u} q\dif\Omega = \frac{4\pi}{3}\mathbf{c},\\
  \TF\!\left(\int_{S^2} \mathbf{u}\mathbf{u}^{\mathsf{T}} q\dif\Omega\right) &= \frac{8\pi}{15}\mathbf{C}.
\end{align*}

The orthogonal projection has the same inner products with all $l\leq 2$ basis functions as the original function, so the coefficients are uniquely determined block-by-block as
\begin{equation*}
  a = \frac{\omega}{4\pi},\quad \mathbf{c} = \frac{3}{4\pi}\mathbf{v},\quad \mathbf{C} = \frac{15}{8\pi}\mathbf{D}.
\end{equation*}

This yields the exact formulas, valid for any integrable angular energy density:
\begin{align*}
  \Pi_1[\ell](\mathbf{u}) &= \frac{\omega}{4\pi} + \frac{3}{4\pi}\mathbf{v}\cdot\mathbf{u},\\[2pt]
  \Pi_2[\ell](\mathbf{u}) &= \frac{\omega}{4\pi} + \frac{3}{4\pi}\mathbf{v}\cdot\mathbf{u}
  + \frac{15}{8\pi}\mathbf{u}^{\mathsf{T}}\mathbf{D}\mathbf{u}.
\end{align*}

This proves that $(\omega,\mathbf{v},\mathbf{D})$ and the full real $l\leq 2$ coefficients are merely two invertible linear coordinate systems.

\subsubsection{Necessity in the Minimal Rotationally Covariant Representation}

Scalars, vectors, and traceless symmetric second-rank tensors respectively carry the $l=0,1,2$ irreducible components of $\operatorname{SO}(3)$. When constructing the corresponding rotationally covariant quantities from the unit direction $\mathbf{u}$ alone, each block can only be---up to a constant factor---$1$, $\mathbf{u}$, and $\TF(\mathbf{u}\mathbf{u}^{\mathsf{T}})$. Imposing first-order homogeneity $F(\lambda\mathbf{x})=\lambda F(\mathbf{x})$ further forces the radial coefficient to be proportional to $r$. Therefore the rigorous meaning of ``must employ the energy--stress tensor'' is: any minimal representation satisfying these requirements must carry information that is invertibly linearly equivalent to it. Storing the nine SH coefficients directly is simply a linear change of basis, not an informationally distinct alternative.

\subsubsection{Incorrect Radial Weighting of the Classical Statistical Second Moment}

The classical mean--covariance Gaussian closure preserves $\EE[\mathbf{X}]$ and $\mathbf{M}_2=\EE[\mathbf{X}\mathbf{X}^{\mathsf{T}}]=\EE[R^2\mathbf{U}\mathbf{U}^{\mathsf{T}}]$, whereas the required stress moment is $\mathbf{S}=\EE[R\mathbf{U}\mathbf{U}^{\mathsf{T}}]$. The radial weights are $R^2$ versus $R$, so $\Tr(\mathbf{M}_2)=\EE[R^2]$ while $\Tr(\mathbf{S})=\EE[R]=\omega$. If all photon energies are scaled by $\lambda R$, then $(\omega,\mathbf{v},\mathbf{D})$ scale uniformly by $\lambda$, while $\mathbf{M}_2$ scales by $\lambda^2$. Hence the classical second moment is neither a first-order homogeneous statistic of radiant energy, nor can its trace substitute for the DC energy.

\subsubsection{A Rigorous Counterexample to Gaussian Identifiability}

Let $\{\mathbf{e}_1,\mathbf{e}_2,\mathbf{e}_3\}$ be the standard orthonormal basis. For any $q_i>0$ with $\sum_i q_i=1$, define the finite photon measure:
\begin{equation*}
  \mu_{\mathbf{q}} = \sum_{i=1}^3 \frac{q_i}{2}
  \bigl(\delta_{\mathbf{e}_i/\sqrt{q_i}} + \delta_{-\mathbf{e}_i/\sqrt{q_i}}\bigr).
\end{equation*}

For all $\mathbf{q}$, $\EE_{\mu_{\mathbf{q}}}[\mathbf{X}]=\mathbf{0}$ and $\EE_{\mu_{\mathbf{q}}}[\mathbf{X}\mathbf{X}^{\mathsf{T}}]=\mathbf{I}$, so they all share the identical classical moment-matched Gaussian $\calN(\mathbf{0},\mathbf{I})$. Yet $\omega_{\mathbf{q}}=\sum_{i=1}^3\sqrt{q_i}$ and $\mathbf{S}_{\mathbf{q}}=\diag(\sqrt{q_1},\sqrt{q_2},\sqrt{q_3})$.

Take $\mathbf{q}^{(0)}=(1/3,1/3,1/3)$ and $\mathbf{q}^{(A)}=(1/2,1/4,1/4)$: their means and classical second moments coincide, but $\omega_0=\sqrt{3}$ and $\omega_A=1+1/\sqrt{2}$. Furthermore let $\mathbf{q}^{(B)}=(1/4,1/2,1/4)$; then $\omega_A=\omega_B$ but $\mathbf{D}_A\neq\mathbf{D}_B$, and their second-order projection difference is $\frac{15}{8\pi}(1/\sqrt{2}-1/2)(u_x^2-u_y^2)\neq 0$. This proves that no universal mapping $(\EE[\mathbf{X}],\Cov(\mathbf{X}))\mapsto(\Pi_1[\ell],\Pi_2[\ell])$ exists. Even with the additional side information $\omega$, no universal mapping $(\omega,\EE[\mathbf{X}],\Cov(\mathbf{X}))\mapsto\Pi_2[\ell]$ exists.

\subsubsection{Insufficiency of the Single-Lobe Spherical Gaussian}

The commonly used axisymmetric single-lobe spherical Gaussian is equivalent to the von Mises--Fisher distribution on $S^2$. Normalized to total energy $\omega$:
\begin{equation*}
  \ell_{\text{SG}}(\mathbf{u})
  = \omega\frac{\lambda}{4\pi\sinh\lambda}\exp(\lambda\,\mathbf{n}\cdot\mathbf{u}),\quad \lambda\geq 0.
\end{equation*}

Let $\mu=\mathbf{n}\cdot\mathbf{u}$. Its normalized first two Legendre moments are
\begin{equation*}
  \rho_1(\lambda)=\coth\lambda-\frac{1}{\lambda},\qquad
  \rho_2(\lambda)=1-\frac{3\rho_1(\lambda)}{\lambda}.
\end{equation*}

Since $\dif\rho_1/\dif\lambda>0$, once the first moment ratio $\rho_1$ is fixed, $\lambda$ is uniquely determined and the second moment $\rho_2$ is forced to a fixed function, incapable of independently carrying the $\zeta$ required by QZH.

A counterexample by equi-moment measures proves this rigorously. Fix any $a\in(-1,1)$: a uniform latitudinal ring measure and a bipolar concentrated measure share the same total energy and the same first moment $\rho_1=a$, yet their second moments are $P_2(a)$ and $1$, respectively. Convolving with smooth narrow kernels on the sphere yields arbitrarily close strictly positive smooth counterexamples.

The single-lobe SG also has two structural limitations: its traceless second moment is necessarily coaxial with the dipole and axisymmetric, unable to express general five-degree-of-freedom traceless stress or biaxial anisotropy; and the single-lobe SG family is not closed under linear mixing---projecting back to a single-lobe model requires nonlinear refitting and does not automatically preserve the second-order stress.


\section{Derivation of the MaxEnt Distribution}

The previous section identified the $1+3+5$ independent moments required to fully preserve the $l\leq 2$ angular energy projection. We now invoke the maximum entropy principle (MaxEnt) under a fixed Lebesgue reference measure $\dif^3\mathbf{x}$ and these moment constraints: among all probability densities satisfying the constraints, choose the one that maximizes differential entropy.

Let $p(\mathbf{x})$ be the probability density of the energy-weighted directional feature $\mathbf{x}\in\RR^3$, with $r=|\mathbf{x}|$. Its differential entropy is
\begin{equation*}
  H[p] = -\int_{\RR^3} p(\mathbf{x})\ln p(\mathbf{x})\dif^3\mathbf{x}.
\end{equation*}

To eliminate the multiplier redundancy caused by the four-dimensional tensor trace constraint, we directly adopt independent statistics:
\begin{equation*}
  s_0(\mathbf{x}) = r,\quad \mathbf{s}_1(\mathbf{x}) = \mathbf{x},\quad
  \mathbf{S}_2(\mathbf{x}) = \TF\!\left(\frac{\mathbf{x}\mathbf{x}^{\mathsf{T}}}{r}\right),
\end{equation*}

with continuous extension $\mathbf{S}_2=\mathbf{0}$ at $r=0$. Given normalized moments $(\bar{\omega},\bar{\mathbf{v}},\bar{\mathbf{D}})$, the constraints are
\begin{align*}
  \int p\dif^3\mathbf{x} &= 1,\qquad
  \int p\,r\dif^3\mathbf{x} = \bar{\omega},\\
  \int p\,\mathbf{x}\dif^3\mathbf{x} &= \bar{\mathbf{v}},\qquad
  \int p\,\mathbf{S}_2\dif^3\mathbf{x} = \bar{\mathbf{D}},\;
  \Tr(\bar{\mathbf{D}})=0.
\end{align*}

Introducing a scalar multiplier $\lambda_0$, a vector multiplier $\boldsymbol{\theta}$, and a traceless symmetric matrix multiplier $\mathbf{A}$, variation of the Lagrangian functional yields:
\begin{align*}
  p(\mathbf{x})
  &=C\exp\!\left(-\beta|\mathbf{x}|+\boldsymbol{\theta}\cdot\mathbf{x}\right.\\
  &\hspace{3.5em}\left.-\frac{\mathbf{x}^{\mathsf{T}}\mathbf{A}\mathbf{x}}{|\mathbf{x}|}\right),\\
  C&=\exp(-1-\lambda_0).
\end{align*}

This form directly uses $1+3+5$ independent constraints, fixes the gauge of the temporal--spatial trace constraint in the four-dimensional packing, and clarifies that the vector term appears only once.

Let $\beta>0$ denote the radial scale and define dimensionless parameters $\mathbf{b}\equiv\boldsymbol{\theta}/\beta$, $\mathbf{Q}\equiv\mathbf{A}/\beta$ ($\Tr(\mathbf{Q})=0$). Introducing spherical coordinates $\mathbf{x}=r\mathbf{u}$ ($r\in[0,\infty)$, $\mathbf{u}\in S^2$), the Cartesian Lebesgue density evaluated at $r\mathbf{u}$ is
\begin{equation*}
  p_{\mathbf X}(r\mathbf{u}) = C\exp(-\beta r g(\mathbf{u})),\qquad
  g(\mathbf{u}) = 1 - \mathbf{b}\cdot\mathbf{u} + \mathbf{u}^{\mathsf{T}}\mathbf{Q}\mathbf{u}.
\end{equation*}
Thus the joint density of $(R,\mathbf U)$ with respect to $\dif r\,\dif\Omega$ includes the spherical-coordinate Jacobian:
\begin{equation*}
  p_{R,\mathbf U}(r,\mathbf{u})
  =p_{\mathbf X}(r\mathbf{u})r^2
  =Cr^2\exp(-\beta r g(\mathbf{u})).
\end{equation*}

In general when $\mathbf{Q}\neq\mathbf{0}$, $\mathbf{b}$ is the natural parameter, while the first moment $\mathbf{v}$ is the gradient of the log-partition function with respect to $\boldsymbol{\theta}$; the two are linked through nonlinear moment matching.

Integrating the joint density over the radial coordinate yields the directional probability marginal. Using
\begin{equation*}
  \int_0^\infty r^2 e^{-a r}\dif r = \frac{2}{a^3},
  \qquad a=\beta g(\mathbf{u})>0,
\end{equation*}
we obtain
\begin{equation*}
  f_{\mathrm{dir}}(\mathbf{u}) = \int_0^\infty p_{\mathbf X}(r\mathbf{u})r^2\dif r
  = \frac{2C}{\beta^3}\frac{1}{g(\mathbf{u})^3}
  = \frac{A}{g(\mathbf{u})^3},
\end{equation*}

where $A$ is determined by $\int_{S^2} f_{\mathrm{dir}}(\mathbf{u})\dif\Omega = 1$. By contrast, the angular energy density contains one additional radial factor because a feature at radius $r$ carries energy $r$:
\begin{align*}
  \ell(\mathbf{u})
  &=\int_0^\infty r\,p_{\mathbf X}(r\mathbf{u})r^2\dif r
    =\frac{6C}{\beta^4}\frac{1}{g(\mathbf{u})^4},\\
  \EE[R\mid\mathbf U=\mathbf{u}]
  &=\frac{\ell(\mathbf{u})}{f_{\mathrm{dir}}(\mathbf{u})}
    =\frac{3}{\beta g(\mathbf{u})}.
\end{align*}
Consequently the unweighted direction marginal is proportional to $g^{-3}$, whereas the energy-weighted moments $(\omega,\mathbf v,\mathbf D)$ and irradiance reconstruction are determined by $\ell\propto g^{-4}$. Sampling $\mathbf U$ from the inferred probability law therefore targets $g^{-3}$; sampling in proportion to the represented angular energy instead targets the normalized $g^{-4}$ density. These two sampling laws are distinct.

\subsection{Existence, Uniqueness, and Parameter Domain}

For interior realizable moments, the log-partition function is strictly convex in the natural parameters; its gradient gives the expectation parameters, so the natural parameters that match the moments are unique after eliminating the gauge mentioned above. The boundary of the realizable domain typically corresponds to weak-limit measures with natural parameters tending to infinity. All finite-parameter formulas in this paper are used within the open domain
\begin{equation*}
  \beta > 0,\quad g_{\min} = \min_{\mathbf{u}\in S^2} g(\mathbf{u}) > 0.
\end{equation*}

\subsection{Partition Function in \texorpdfstring{$\RR^3$}{R3}}

Substituting the simplified parameters, the unnormalized spatial probability density is
\begin{equation*}
  p(\mathbf{x}) = \frac{1}{Z}\exp\!\left(-\beta|\mathbf{x}|
  \Bigl(1-\mathbf{b}\cdot\frac{\mathbf{x}}{|\mathbf{x}|}
  +\frac{\mathbf{x}^{\mathsf{T}}\mathbf{Q}\mathbf{x}}{|\mathbf{x}|^2}\Bigr)\right).
\end{equation*}

Integrating over $\RR^3$, using spherical coordinates and separating radial and angular integrals:
\begin{equation*}
  Z = \int_{S^2}\!\left(\int_0^\infty r^2\exp(-r\beta g(\mathbf{u}))\dif r\right)\!\dif\Omega
  = \frac{2}{\beta^3}\int_{S^2}\frac{1}{g(\mathbf{u})^3}\dif\Omega.
\end{equation*}

Defining the purely angular partition function $Z_\Omega = \int_{S^2} g(\mathbf{u})^{-3}\dif\Omega$, the spatial partition function is $Z_3 = (2Z_\Omega)/\beta^3$. When $\mathbf{Q}=\mathbf{0}$ and $\mathbf{b}=\kappa\mathbf{n}$, $Z_\Omega = (4\pi)/(1-\kappa^2)^2$.

\subsection{Degree-of-Freedom Accounting}

We must distinguish the normalization constant of the probability distribution from the total energy scale:
\begin{itemize}
  \item The normalized $\RR^3$ distribution is parameterized by $\beta$ (1 DOF), $\mathbf{b}$ (3 DOF), and $\mathbf{Q}$ (5 DOF), totaling 9 DOF;
  \item The normalized spherical marginal distribution eliminates $\beta$, leaving only the 8 shape degrees of freedom in $\mathbf{b}$ and $\mathbf{Q}$;
  \item If the spherical representation must also carry the total radiant energy $\omega$, we regain $1+3+5=9$ storage DOF.
\end{itemize}

The normalization coefficient $A = I(\mathbf{b},\mathbf{Q})^{-1}$ is uniquely determined by the shape parameters, is not an independent DOF, and is not equal to $\omega$. The natural parameters and empirical moments are linked through the gradient equations of the partition function via nonlinear moment matching; in general when $\mathbf{Q}\neq\mathbf{0}$, $\mathbf{b}$ and $\mathbf{Q}$ cannot be equated component-wise with $\bar{T}^{0j}$ and $\TF(\bar{T}^{jk})$.


\section{Properties of the Complete Second-Order Distribution}

This section studies the general traceless symmetric tensor without special assumptions about its eigenvalues, eigenvectors, or their relative orientation with respect to the dipole direction. Define
\begin{equation*}
  g(\mathbf{u}) = 1 - \mathbf{b}\cdot\mathbf{u} + \mathbf{u}^{\mathsf{T}}\mathbf{Q}\mathbf{u},\quad \Tr(\mathbf{Q})=0.
\end{equation*}

The normalized distribution on $\RR^3$ and its spherical marginal are respectively
\begin{align*}
  p(\mathbf{x}) &= \frac{1}{Z}\exp(-\beta|\mathbf{x}|g(\hat{\mathbf{x}})),\\
  f(\mathbf{u}) &= \frac{g(\mathbf{u})^{-3}}{I(\mathbf{b},\mathbf{Q})},\quad
  I(\mathbf{b},\mathbf{Q}) = \int_{S^2} g(\mathbf{u})^{-3}\dif\Omega,
\end{align*}
with the full partition function $Z(\beta,\mathbf{b},\mathbf{Q}) = \frac{2}{\beta^3} I(\mathbf{b},\mathbf{Q})$.

\subsection{Normalizability Condition}

The necessary and sufficient condition for normalizability is
\begin{equation*}
  g_{\min}=\min_{\mathbf{u}\in S^2}g(\mathbf{u})>0.
\end{equation*}
A convenient sufficient (but not necessary) condition is $|\mathbf{b}|+\norm{\mathbf{Q}}_{\mathrm{op}}<1$. When $\mathbf{b}=\mathbf{0}$, the necessary and sufficient condition reduces to $1+\lambda_{\min}(\mathbf{Q})>0$.

\subsection{Taylor Expansion and Error Bound}

Define the angular perturbation $h(\mathbf{u})=\mathbf{b}\cdot\mathbf{u}-\mathbf{u}^{\mathsf{T}}\mathbf{Q}\mathbf{u}$, so $g=1-h$. If $\rho=\max_{\mathbf{u}\in S^2}|h(\mathbf{u})|<1$, the binomial series converges uniformly and absolutely on the entire sphere:
\begin{equation*}
  g(\mathbf{u})^{-3} = (1-h)^{-3} = \sum_{n=0}^\infty \frac{(n+1)(n+2)}{2} h^n.
\end{equation*}

Truncating to second order gives $g^{-3}=1+3h+6h^2+R_2(h)$, with the exact remainder
\begin{equation*}
  R_2(h) = \frac{h^3(10-15h+6h^2)}{(1-h)^3},
\end{equation*}
and $\sup_{|h|\leq\rho}|R_2(h)| = \rho^3(10-15\rho+6\rho^2)/(1-\rho)^3$, with equality attained at $h=\rho$.

\subsection{Second-Order Approximation of the Partition Function}

Using spherical symmetry integration formulas:
\begin{align*}
  I(\mathbf{b},\mathbf{Q}) &= 4\pi\Bigl(1 + 2|\mathbf{b}|^2 + \frac{4}{5}\Tr(\mathbf{Q}^2)\Bigr) + O(\rho^3),\\
  Z &= \frac{8\pi}{\beta^3}\Bigl(1 + 2|\mathbf{b}|^2 + \frac{4}{5}\Tr(\mathbf{Q}^2)\Bigr) + O(\rho^3).
\end{align*}

The second-order parameter expansion of the normalized spherical distribution is
\begin{equation*}
  f(\mathbf{u}) = \frac{1}{4\pi}\Bigl(1 + 3h + 6h^2 - 2|\mathbf{b}|^2 - \frac{4}{5}\Tr(\mathbf{Q}^2)\Bigr) + O(\rho^3).
\end{equation*}

\subsection{Local Relationship with SH2}

When only first-order parameter perturbations are retained,
\begin{equation*}
  f(\mathbf{u}) = \frac{1}{4\pi}\bigl(1 + 3\mathbf{b}\cdot\mathbf{u} - 3\mathbf{u}^{\mathsf{T}}\mathbf{Q}\mathbf{u}\bigr) + O(\rho^2),
\end{equation*}
whose right-hand side lies exactly in the SH2 space. Hence the MaxEnt family near the uniform distribution has SH2 as its tangent space, with first-order parameter mapping $c_0=1/4\pi$, $\mathbf{c}_1=3\mathbf{b}/4\pi$, $\mathbf{C}_2=-3\mathbf{Q}/4\pi$. This is, however, only a local first-order equivalence, not a global equivalence between the full distribution and SH2: $h$ contains both first- and second-order spherical harmonic components, $h^n$ contains components up to degree $2n$, and the full MaxEnt distribution generally contains infinitely many spherical harmonic bands.

\subsection{SH2 Projection of the Second-Order Truncation}

Let $\Pi_2$ denote the orthogonal projection onto the $l\leq 2$ spherical harmonic space. The low-frequency projection of the quadratic term is
\begin{multline*}
  \Pi_2[h^2] = \frac{|\mathbf{b}|^2}{3} + \frac{2}{15}\Tr(\mathbf{Q}^2)
  - \frac{4}{5}\mathbf{Q}\mathbf{b}\cdot\mathbf{u}\\
  + \mathbf{u}^{\mathsf{T}}\Bigl[\TF(\mathbf{b}\mathbf{b}^{\mathsf{T}}) + \frac{4}{7}\TF(\mathbf{Q}^2)\Bigr]\mathbf{u}.
\end{multline*}

Substituting into the normalized expansion, the constant correction cancels exactly, yielding
\begin{equation*}
  \Pi_2[f](\mathbf{u}) = \frac{1}{4\pi}\Bigl[1 + \bigl(3\mathbf{b} - \tfrac{24}{5}\mathbf{Q}\mathbf{b}\bigr)\cdot\mathbf{u}
  + \mathbf{u}^{\mathsf{T}}\mathbf{S}\mathbf{u}\Bigr] + O(\rho^3),
\end{equation*}
where $\mathbf{S} = -3\mathbf{Q} + 6\TF(\mathbf{b}\mathbf{b}^{\mathsf{T}}) + \frac{24}{7}\TF(\mathbf{Q}^2)$. This gives the second-order nonlinear mapping from MaxEnt parameters to SH2 coefficients.

\subsection{Positivity and High-Frequency Reconstruction}

As long as $g_{\min}>0$, the full distribution strictly satisfies $f(\mathbf{u})>0$ and is smooth over the entire sphere. In contrast, a general SH2 function carries no automatic positivity guarantee. The full distribution generates all higher-frequency bands from the low-order sufficient statistics through the nonlinear function $g^{-3}$, and can therefore form sharper angular structures than SH2 while maintaining a finite parameter count.

\subsection{Peaks and Stationary Points}

Distribution peaks are equivalent to minima of $g$ on the sphere. Stationary points satisfy $(\mathbf{Q}-\lambda\mathbf{I})\mathbf{u}=\frac{1}{2}\mathbf{b}$, showing that the peak direction is generally neither parallel to $\mathbf{b}$ nor equal to any eigenvector of $\mathbf{Q}$.

\subsection{Rotational Covariance and Antipodal Symmetry}

For any $\mathbf{R}\in\operatorname{SO}(3)$, the distribution is covariant when parameters transform as $\mathbf{b}\to\mathbf{R}\mathbf{b}$, $\mathbf{Q}\to\mathbf{R}\mathbf{Q}\mathbf{R}^{\mathsf{T}}$. The antipodal density ratio is
\begin{equation*}
  \frac{f(\mathbf{u})}{f(-\mathbf{u})}
  = \left(\frac{1+\mathbf{b}\cdot\mathbf{u}+\mathbf{u}^{\mathsf{T}}\mathbf{Q}\mathbf{u}}
  {1-\mathbf{b}\cdot\mathbf{u}+\mathbf{u}^{\mathsf{T}}\mathbf{Q}\mathbf{u}}\right)^3,
\end{equation*}
showing that $\mathbf{Q}$ itself has antipodal symmetry, while all antipodal asymmetry is introduced by the dipole parameter $\mathbf{b}$.

\subsection{Radial Conditional Distribution}

Conditioned on direction $\mathbf{u}$, the radial density is Gamma-distributed:
\begin{equation*}
  p(r\mid\mathbf{u}) = \frac{(\beta g(\mathbf{u}))^3}{2} r^2 \exp(-\beta g(\mathbf{u}) r),\quad r\geq 0,
\end{equation*}
hence $\EE[r\mid\mathbf{u}] = 3/(\beta g(\mathbf{u}))$ and $\Var[r\mid\mathbf{u}] = 3/(\beta^2 g(\mathbf{u})^2)$.

\subsection{DOF and Distinction from SH2}

Although SH2 and this distribution share the same finite parameter count, they serve different roles: SH2 is a nine-dimensional linear function space supporting linear accumulation and filtering, but without automatic positivity; MaxEnt is a nine-dimensional nonlinear distribution family whose parameters are generally nonlinearly related to the moments, but which is strictly positive throughout the normalizable domain and generates constrained high-frequency content through its nonlinear form.


\section{First-Order Reduction: Analytic Closure of the Pure-Dipole Model}

The pure dipole model sets $\mathbf{Q}=\mathbf{0}$, retaining only total energy and the first directional moment. It targets scenarios with constrained storage or compute budgets, or where second-order anisotropy is weak.

The aggregate encoding then degenerates to an augmented vector $L = (\mathbf{v},\omega) \in \RR^3\times\RR_{\geq 0}$. This state space forms the augmented convex cone $\calC = \{(\mathbf{v},\omega)\mid\omega\geq|\mathbf{v}|\}$. Substituting $\mathbf{Q}=\mathbf{0}$, the macroscopic anisotropy is defined as $\rho=|\mathbf{v}|/\omega$. When $\mathbf{v}\neq\mathbf{0}$, let $\mathbf{n}=\mathbf{v}/|\mathbf{v}|$; then $\mathbf{b}=\kappa\mathbf{n}$. Within the normalizable domain $\beta>0$, $0\leq\kappa<1$:
\begin{equation*}
  Z_3 = \frac{8\pi}{\beta^3(1-\kappa^2)^2},\qquad
  f_{\text{dir}}(\mathbf{u}) = \frac{(1-\kappa^2)^2}{4\pi(1-\kappa\mathbf{n}\cdot\mathbf{u})^3}.
\end{equation*}

Conditioned on direction, the radius follows a Gamma distribution with rate $\beta(1-\kappa\mathbf{n}\cdot\mathbf{u})$ and shape $3$. The angular energy density is
\begin{equation*}
  \ell(\mathbf{u}) = \omega\frac{3(1-\kappa^2)^3}{4\pi(3+\kappa^2)}\frac{1}{(1-\kappa\mathbf{n}\cdot\mathbf{u})^4},\quad
  \int_{S^2}\ell\dif\Omega = \omega.
\end{equation*}

Moment matching gives:
\begin{equation*}
  \rho = \frac{|\mathbf{v}|}{\omega} = \frac{4\kappa}{3+\kappa^2},\qquad
  \kappa = \frac{3\rho}{2+\sqrt{4-3\rho^2}}.
\end{equation*}
The latter form avoids cancellation error near $\rho\approx 0$ through denominator rationalization.

\subsection{Analytic Closure of the Joint Covariance in Augmented Space}

Under first-order reduction, the spatial covariance matrix has perfect axisymmetric geometry:
\begin{equation*}
  \Cov(\mathbf{X}) = \sigma_{\perp}^2(\mathbf{I}-\hat{\mathbf{v}}\hat{\mathbf{v}}^{\mathsf{T}}) + \sigma_{\parallel}^2\hat{\mathbf{v}}\hat{\mathbf{v}}^{\mathsf{T}},
\end{equation*}
where $\sigma_{\perp}^2 = 4\omega^2(1-\kappa^2)/(3+\kappa^2)^2$ and $\sigma_{\parallel}^2 = 4\omega^2(1+\kappa^2)/(3+\kappa^2)^2$. The scalar variance is
\begin{equation*}
  \Var_{\text{scalar}}(\mathbf{X}) = \frac{2\omega^2 + \omega\sqrt{4\omega^2-3|\mathbf{v}|^2}}{3} - \frac{1}{2}|\mathbf{v}|^2,
\end{equation*}
ranging over $[\frac{1}{2}\omega^2,\frac{4}{3}\omega^2]$. The full $4\times 4$ covariance closure for the augmented variable $\mathbf{Y}=(\mathbf{X},R)$ also exists (cross-covariance $8\kappa\omega^2\hat{\mathbf{v}}/(3+\kappa^2)^2$, radial variance $\omega^2(3+6\kappa^2-\kappa^4)/(3+\kappa^2)^2$).

\textit{Note: the above variances are those of the MaxEnt distribution inferred from moment estimates, not the sample variances. Distributional variances cannot serve as substitutes for sample variances.}

\subsection{Closed-Form Diffuse Lighting Reconstruction}

Let $\mu_0=\hat{\mathbf{v}}\cdot\mathbf{n}$, and define $d=1-\kappa^2+\kappa^2\mu_0^2$ and $N_S=3+6\kappa^2(-1+2\mu_0^2)+\kappa^4(3-12\mu_0^2+8\mu_0^4)$. The exact closed form for the raw shape response $e_{\text{raw}}=E_{\text{raw}}/\omega$ is
\begin{equation*}
  e_{\text{raw}}(\mu_0,\kappa) = \frac{N_S + 8\kappa\mu_0 d^{3/2}}{4(3+\kappa^2)d^{3/2}},\qquad
  e(\mu_0,\kappa)=\frac{e_{\text{raw}}(\mu_0,\kappa)}{\pi}.
\end{equation*}

The symmetric and antisymmetric parts are respectively $e_{\text{raw},S} = N_S/(4(3+\kappa^2)d^{3/2})$ and $e_{\text{raw},A} = 2\kappa\mu_0/(3+\kappa^2) = \rho\mu_0/2$.

All boundary cases: $\kappa\to 0$ gives $1/4$; $\kappa\to 1$ gives $\max(0,\mu_0)$; $\mu_0=1$ gives $(1+\kappa)^3(3-\kappa)/(4(3+\kappa^2))$; $\mu_0=0$ gives $3\sqrt{1-\kappa^2}/(4(3+\kappa^2))$.

The following HLSL implementation corresponds term-by-term with the exact closed form:
\begin{lstlisting}[language=C, caption={Pure dipole normalized-cosine diffuse reconstruction}]
float ReconstructDiffuse(float3 v, float omega, float3 n)
{
    if (omega <= 1e-6f) return 0.0f;
    float lenV = length(v);
    if (lenV <= 1e-6f) return (0.25f / PI) * omega;

    float rho = clamp(lenV / omega, 0.0f, 1.0f);
    float kappa = (3.0f * rho) /
        (2.0f + sqrt(max(0.0f, 4.0f - 3.0f * rho * rho)));
    float mu = clamp(dot(v / lenV, n), -1.0f, 1.0f);

    if (1.0f - kappa <= 1e-5f)
        return (omega / PI) * max(mu, 0.0f);

    float k2 = kappa * kappa, mu2 = mu * mu;
    float d = max(1e-12f, 1.0f - k2 + k2 * mu2);
    float d32 = d * sqrt(d);
    float nSym = 3.0f + 6.0f * k2 * (-1.0f + 2.0f * mu2)
        + k2 * k2 * (3.0f - 12.0f * mu2 + 8.0f * mu2 * mu2);
    float e = (nSym + 8.0f * kappa * mu * d32)
        / (4.0f * (3.0f + k2) * d32);
    return (omega / PI) * max(e, 0.0f);
}
\end{lstlisting}

The theoretical closed form is non-negative for $0\leq\kappa<1$; the trailing \texttt{max} only suppresses tiny negatives from floating-point rounding. Under the normalized cosine convention the isotropic output is $\omega/(4\pi)$; for the raw hemispherical cosine integral, omit the \texttt{/ PI}.

\subsection{Path Guiding: Analytic Zero-Rejection Inverse Sampling}

The pure dipole directional probability $g^{-3}$ admits an exact zero-rejection inverse CDF: let $\mu=\hat{\mathbf{v}}\cdot\mathbf{u}$, and generate a direction from uniform random numbers $r_1,r_2\in[0,1)$ with $\phi=2\pi r_2$. The analytic inverse mapping for the polar offset is $\mu = \bigl(1-1/\sqrt{A+Br_1}\bigr)/\kappa$, where $A=(1/(1+\kappa))^2$ and $B=(1/(1-\kappa))^2-A$. The continuous limit $\kappa\to 0$ is $\mu=2r_1-1$ (uniform spherical sampling); as $\kappa\to 1$ the directional probability progressively concentrates toward the positive principal axis.

Note the sampling target distinction: the square-root inverse formula above samples the directional probability $f\propto g^{-3}$. Diffuse reconstruction uses $\ell\propto g^{-4}$, which corresponds to a cube-root inverse and cannot be reused directly.

\subsection{Information Geometry and State Metrics}

\subsubsection{Closed-Form Bures--Wasserstein Distance}

Construct moment-matched Gaussian proxies from the first moment and covariance of each MaxEnt state. The squared Bures--Wasserstein distance between two proxies~\cite{Bhatia2019Bures} is:
\begin{multline*}
  d_B^2(L_1,L_2) = \|\mathbf{v}_1-\mathbf{v}_2\|^2 + \Tr(\boldsymbol{\Sigma}_1) + \Tr(\boldsymbol{\Sigma}_2)\\
  - 2\Tr\bigl((\boldsymbol{\Sigma}_1^{1/2}\boldsymbol{\Sigma}_2\boldsymbol{\Sigma}_1^{1/2})^{1/2}\bigr).
\end{multline*}

Since the spatial covariance of this system has an exact axisymmetric form, the cross-trace operation is strictly closed within the two-dimensional subspace spanned by $\mathbf{v}_1,\mathbf{v}_2$. When $|\mathbf{v}_1||\mathbf{v}_2|>0$, let $c$ be the cosine of the angle between them, and $\sigma_\perp,\sigma_\parallel$ be the perpendicular and parallel eigen-standard-deviations. Define
\begin{align*}
  A &= \sigma_{\perp,1}\sigma_{\parallel,2}
     + \sigma_{\perp,2}\sigma_{\parallel,1},\\
  B_i &= \sigma_{\parallel,i}^2-\sigma_{\perp,i}^2,
\end{align*}
then the cross-trace can be written compactly as
\begin{equation*}
  \Tr_{\mathrm{cross}}
  = \sigma_{\perp,1}\sigma_{\perp,2}
  + \sqrt{A^2+c^2B_1B_2}.
\end{equation*}

This distance is exact for the Gaussian proxies, but is not the exact Wasserstein distance between two non-Gaussian MaxEnt distributions.

\subsubsection{Natural Parameters and Weighted Jeffreys Divergence}

The analytic mapping for the first-order reduced system is
\begin{equation*}
  \beta = \frac{3+\kappa^2}{\omega(1-\kappa^2)},\qquad
  \boldsymbol{\theta} = \frac{(3+\kappa^2)^2}{4\omega^2(1-\kappa^2)}\mathbf{v}.
\end{equation*}

The Jeffreys divergence (exactly symmetrized KL divergence) between two pure dipole MaxEnt states is
\begin{equation*}
  D_J(L_1,L_2) = \beta_1\omega_2 + \beta_2\omega_1 - \boldsymbol{\theta}_1\cdot\mathbf{v}_2 - \boldsymbol{\theta}_2\cdot\mathbf{v}_1 - 6.
\end{equation*}


\section{Coaxial Five-Parameter MaxEnt Subfamily}

The full 9-DOF model allows the dipole natural parameter and the stress tensor to have an arbitrary relative orientation. Requiring them to share a single axis yields a closed subfamily that still retains second-order information but needs only five degrees of freedom. Choose a unit axis $\mathbf{n}$ and define
\begin{equation*}
  \mathbf{b} = \kappa\mathbf{n},\qquad
  \mathbf{Q}(\alpha) = \frac{\alpha}{2}(3\mathbf{n}\mathbf{n}^{\mathsf{T}} - \mathbf{I}),\quad |\mathbf{n}|=1.
\end{equation*}

$\mathbf{Q}(\alpha)$ is symmetric and traceless, with eigenvalues $\alpha$ (along the axis) and $-\alpha/2$ (transverse). Let $\mu=\mathbf{n}\cdot\mathbf{u}$ and $P_2(\mu)=\frac{1}{2}(3\mu^2-1)$; then $\mathbf{u}^{\mathsf{T}}\mathbf{Q}\mathbf{u}=\alpha P_2(\mu)$. The general 9-DOF distribution thus reduces to
\begin{equation*}
  g(\mu) = 1 - \kappa\mu + \alpha P_2(\mu),\quad
  p(\mathbf{x}) = \frac{1}{Z}\exp(-\beta|\mathbf{x}|g(\mu)).
\end{equation*}

The natural parameters are $\beta$, $\kappa$, $\alpha$, and the axis $\mathbf{n}\in S^2$, totaling $1+1+1+2=5$ DOF. The corresponding linear moments can be written as total energy $\omega$, axial momentum magnitude $v_\parallel$, axial second moment $\zeta$, and axis $\mathbf{n}$:
\begin{equation*}
  \mathbf{v} = v_\parallel\mathbf{n},\quad
  \zeta = \EE[rP_2(\mu)],\quad
  \EE[r\mathbf{u}\mathbf{u}^{\mathsf{T}}] = \frac{\omega}{3}\mathbf{I} + \frac{\zeta}{3}(3\mathbf{n}\mathbf{n}^{\mathsf{T}}-\mathbf{I}).
\end{equation*}

When $|\mathbf{v}|>0$, the axis can be recovered from $\mathbf{v}/|\mathbf{v}|$, and the state is equivalent to $(\omega,\mathbf{v},\zeta)$, totaling $1+3+1=5$ scalars. In the zero-dipole but second-order-anisotropic branch, $\zeta$ alone cannot determine the unoriented stress axis; this degenerate branch must retain the stress axis or recover it from the full stress tensor. Adding two five-parameter states with different axes generally leaves the subfamily; practical filtering should first accumulate in the linear 9-DOF tensor, then project back to the coaxial subfamily.

\subsection{Moment-Space Mixing and Coaxial Projection}

The forward mapping from dimensionless state $(\omega,\rho,\eta,\mathbf{n})$ to linear tensor coordinates $(\omega,\mathbf{v},\mathbf{T})$ is
\begin{equation*}
  \mathbf{v} = \rho\omega\mathbf{n},\qquad
  \mathbf{T} = \frac{\eta\omega}{3}(3\mathbf{n}\mathbf{n}^{\mathsf{T}} - \mathbf{I}).
\end{equation*}

Given linear coordinates, the inverse recovery: when $|\mathbf{v}|>0$, axis $\mathbf{n}=\mathbf{v}/|\mathbf{v}|$ and $\zeta=\frac{3}{2}\mathbf{n}^{\mathsf{T}}\mathbf{T}\mathbf{n}$; when $\mathbf{v}=\mathbf{0}$ and $\mathbf{T}\neq\mathbf{0}$, take the eigenvector of $\mathbf{T}$ with the largest absolute eigenvalue as the unoriented axis. Weighted mixing first accumulates linearly in tensor space, then projects back to the subfamily using the inverse transform. The realizable domain satisfies $P_2(\rho)\leq\eta\leq 1$ (zero-dipole branch: $-1/2\leq\eta\leq 1$).

\subsection{Exact Normalizability Domain}

Writing $g$ as a quadratic polynomial $g(\mu)=A\mu^2+B\mu+C$ with $A=3\alpha/2$, $B=-\kappa$, $C=1-\alpha/2$, and taking $\kappa\geq 0$ (by branch symmetry), the necessary and sufficient normalizability condition is:
\begin{equation*}
  \begin{cases}
    1+\alpha-\kappa > 0 & \alpha\leq 0\ \text{or}\ \kappa\geq 3\alpha,\\[4pt]
    1-\dfrac{\alpha}{2}-\dfrac{\kappa^2}{6\alpha} > 0 & \alpha>0\ \text{and}\ 0\leq\kappa\leq 3\alpha.
  \end{cases}
\end{equation*}

\subsection{Principal Integral Closed Form and Partition Function}

Define the angular principal integral $Q(\kappa,\alpha)=\int_{-1}^1 g(\mu)^{-3}\dif\mu$, and let $g_{\pm}=1+\alpha\mp\kappa$, $\Delta=3\alpha(2-\alpha)-\kappa^2$. When $\Delta>0$,
\begin{align*}
  Q(\kappa,\alpha)
  &= \frac{N_3(\kappa,\alpha)}{g_{+}^2g_{-}^2\Delta^2}\\
  &\quad + \frac{27\alpha^2}{\Delta^{5/2}}
  \left(\arctan\frac{3\alpha+\kappa}{\sqrt{\Delta}}
  + \arctan\frac{3\alpha-\kappa}{\sqrt{\Delta}}\right).
\end{align*}
The polynomial in the rational term is
\begin{align*}
  N_3(\kappa,\alpha)={}&18\alpha^5+81\alpha^4+108\alpha^3+45\alpha^2\\
  &-\left(42\alpha^3+39\alpha^2+24\alpha\right)\kappa^2\\
  &+\left(8\alpha+2\right)\kappa^4.
\end{align*}
For $\Delta<0$, let $s=\sqrt{-\Delta}$. The corresponding real logarithmic branch is
\begin{align*}
  Q(\kappa,\alpha)
  ={}&\frac{N_3(\kappa,\alpha)}{g_{+}^2g_{-}^2\Delta^2}\\
  &+\frac{27\alpha^2}{2s^5}
  \ln\left|\frac{2\alpha-1-s}{2\alpha-1+s}\right|.
\end{align*}
This reduced logarithmic form remains well defined in the pure-dipole limit $\alpha\to0$. At $\Delta=0$ the two terms of either non-degenerate branch must not be evaluated separately. Direct integration of $g(\mu)=(3\alpha\mu-\kappa)^2/(6\alpha)$ gives, for $\alpha\neq0$ in the normalizable domain,
\begin{equation*}
  Q\big|_{\Delta=0}
  =-\frac{72\alpha^2}{5}
  \left[\frac{1}{(3\alpha-\kappa)^5}
  +\frac{1}{(3\alpha+\kappa)^5}\right],
\end{equation*}
with $Q(0,0)=2$. Equivalently, this is the joint continuous limit of the $\Delta>0$ and $\Delta<0$ expressions. The partition function is $Z(\beta,\kappa,\alpha) = (4\pi/\beta^3)Q(\kappa,\alpha)$, and all derivatives of $\ln Z$ with respect to the natural parameters are elementary closed forms.

\subsection{Analytic Closure of Partition Function and Five Moments}

Logarithmic derivatives give the angular first moment and radial coupling:
\begin{align*}
  \partial_\kappa\ln Q &= \beta v_\parallel,\qquad
  \partial_\alpha\ln Q = -\beta\zeta,\\
  \omega &= \frac{3}{\beta} + \kappa v_\parallel - \alpha\zeta.
\end{align*}

Hence the forward mapping:
\begin{align*}
  \beta\omega &= 3 + \kappa\partial_\kappa\ln Q + \alpha\partial_\alpha\ln Q,\\
  \rho &= \frac{\partial_\kappa\ln Q}{3+\kappa\partial_\kappa\ln Q+\alpha\partial_\alpha\ln Q},\\
  \eta &= -\frac{\partial_\alpha\ln Q}{3+\kappa\partial_\kappa\ln Q+\alpha\partial_\alpha\ln Q}.
\end{align*}

The angular energy density $\ell(\mu)\propto g(\mu)^{-4}$ also requires normalization through $W(\kappa,\alpha)=\int_{-1}^1 g(\mu)^{-4}\dif\mu = (\beta\omega/3)Q$. The inverse mapping generally involves both inverse-trigonometric or logarithmic terms; in practice a two-dimensional Newton method can be used within the exact normalizable domain, with the analytic Jacobian obtained directly from the covariance closed form.

\subsection{Spatial and Augmented-Space Covariance}

The Hessian of the log-partition function yields the Fisher information blocks. Define $I_{\kappa\kappa}=\partial_\kappa^2\ln Q$, $I_{\kappa\alpha}=\partial_\kappa\partial_\alpha\ln Q$, $I_{\alpha\alpha}=\partial_\alpha^2\ln Q$ (all three are elementary closed forms). The complete expressions for angular covariance and radial correction give the $4\times 4$ covariance closure of the augmented variable $\mathbf{Y}=(\mathbf{X},R)$. In the limit $\alpha\to 0$ all quantities degenerate to the parallel/perpendicular variances of the pure dipole chapter---strictly consistent.

\subsection{Normal-Axis One-Dimensional Form for Diffuse Lighting Reconstruction}

The angular energy density is $\ell(\mathbf{u}) = (\omega/2\pi W)g(\mu)^{-4}$. Taking the surface normal $\mathbf{n}_N$ as the polar axis, let $m=\mathbf{n}\cdot\mathbf{n}_N$ and $z=\cos\theta$. The raw shape response is
\begin{align*}
  e_{\text{raw}}(m,\kappa,\alpha) &= \frac{1}{2\pi W}\int_0^1 z H(z)\dif z,\\
  H(z) &= \int_0^{2\pi}\frac{\dif\phi}{g(x+y\cos\phi)^4}.
\end{align*}

\subsubsection{Quadratic Cosine Denominator and Palindromic Quartic}

Let $c_\phi=\cos\phi$; then $g(x+yc_\phi)=ac_\phi^2+bc_\phi+c$, where $a=3\alpha y^2/2$, $b=y(3\alpha x-\kappa)$, $c=1-\alpha/2-\kappa x+3\alpha x^2/2$. Changing variables on the unit circle $w=e^{i\phi}$ and defining the palindromic quartic $P_z(w)=aw^4+2bw^3+(2a+4c)w^2+2bw+a$, we have the identity $g=P_z(w)/(4w^2)$. The inner integral becomes a rational-function contour integral on the unit circle: $H(z)=\frac{256}{i}\oint_{|w|=1}\frac{w^7}{P_z(w)^4}\dif w$.

\subsubsection{Purely Real Algebraic Elimination of the Residue Sum}

Let
\begin{equation*}
  r=\sqrt{(a+c)^2-b^2},\qquad
  q=\sqrt{(a+c+r)(c-a+r)}.
\end{equation*}
Adding the two fourth-order pole residues inside the unit circle and eliminating roots using the palindromic relation yields
\begin{equation*}
  H(z) = -\frac{\pi q P(a,c,r)}{(c-a+r)^4 r^7},
\end{equation*}
where
\begin{align*}
  P(a,c,r)
  &= 5 (a - c)^3 (a + c)^3
   - 5 (a - c)^2 (a + c)^2 (4 a + 3 c) r \\
  &\quad + 2 (a - c) (a + c)
   (13 a^2 + 24 a c + 6 c^2) r^2 \\
  &\quad + (-4 a^3 - 45 a^2 c - 32 a c^2 + 4 c^3) r^3 \\
  &\quad - (4 a - 3 c) (4 a + 3 c) r^4
   + (4 a + 3 c) r^5.
\end{align*}
This is the final runtime form of the inner angular integral: only scalar addition, multiplication, division, and two real square roots are needed; no complex roots or logarithms. Substituting back, the raw shape response reduces to the one-dimensional integral
\begin{equation*}
  e_{\text{raw}}(m,\kappa,\alpha)
  = -\frac{1}{2W}\int_0^1
  \frac{z\,q(z)\,P(a(z),c(z),r(z))}
       {(c(z)-a(z)+r(z))^4\,r(z)^7}\dif z.
\end{equation*}
The result is derived by the unit-circle residue method and can also be obtained by explicit differentiation of an independent real integral identity three times.

Continuity limits: $a=0$ (including $\alpha=0$) gives $H(z)=\pi c(2c^2+3b^2)/(c^2-b^2)^{7/2}$, the azimuthal closed form of the four-parameter chapter; $b=0$ and $a=0$ reduces to $H(z)=2\pi/c^4$. The theoretical formula requires no special branches; for floating-point stability it can be evaluated in logarithmic-derivative form involving only rational operations and two \texttt{sqrt} calls, without invoking \texttt{log}.

The antisymmetric part requires no residue calculation: for any normalized angular energy density, $e_{\text{raw}}(m)-e_{\text{raw}}(-m)=\rho m$. The second-order parameter only alters the symmetric part of the response and the mapping from natural parameters to $\rho$.

\subsection{Offline Response Table Generation and Shader Reconstruction}

The outer one-dimensional integral is a controlled numerical algorithm requiring per-node evaluation. For real-time pipelines, a more suitable approach is to treat the one-dimensional form as an offline ground-truth generator, build a response table directly on the stored moment parameters, and let the shader perform a single hardware trilinear sample.

\subsubsection{Offline Generation and Table Compression}

The stored moments $(\rho,\eta)$ and natural parameters $(\kappa,\alpha)$ are linked through the moment-matching equations, whose inverse generally lacks an elementary closed form. During offline table construction a two-dimensional Newton method solves for the natural parameters, after which the purely algebraic formula evaluates the response. The antisymmetric part is recovered analytically, so the table only needs to store the back response $B_{\text{raw}}(\rho,q,u)=e_{\text{raw}}(-u)$ ($u=|m|$). At runtime a single trilinear sample recovers: $e_{\text{raw}}(m)=B_{\text{raw}}(\rho,q,|m|)+\rho\max(m,0)$.

\subsubsection{Moment-Domain Coordinates}

Using the realizable interval $\eta\in[P_2(\rho),1]$, define the normalized second-order coordinate
\begin{equation*}
  q = \begin{cases}
    \dfrac{\eta-P_2(\rho)}{1-P_2(\rho)} & \rho<1,\\[8pt]
    1 & \rho=1.
  \end{cases}
\end{equation*}

The table coordinates are $(\rho,q,u)\in[0,1]^3$. The $\rho\to 1$ and $q\to 0$ edges require nonlinear grids to allocate more nodes near the boundaries.

\subsubsection{Shader Reconstruction}

\begin{lstlisting}[language=C, caption={Coaxial five-parameter MaxEnt diffuse reconstruction (LUT sample)}]
float WarpRho(float rho) {
    return 1.0f - pow(max(1.0f - saturate(rho), 0.0f), 1.0f/3.0f);
}
float WarpQ(float q) {
    return 0.5f * (1.0f - cos(PI * saturate(q)));
}

float ReconstructAxisAligned(float omega, float3 v,
    float zeta, float3 stressAxis, float3 normal, float lutSize)
{
    if (omega <= 1e-8f) return 0.0f;
    float lenV = length(v);
    float rho = saturate(lenV / omega);
    float3 axis = lenV > 1e-7f ? v / lenV : stressAxis;
    float m = clamp(dot(axis, normal), -1.0f, 1.0f);
    float eta = zeta / omega;
    float etaMin = 0.5f * (3.0f * rho * rho - 1.0f);
    float q = saturate((eta - etaMin) / max(1.0f - etaMin, 1e-7f));

    float3 gridCoord = float3(abs(m), WarpQ(q), WarpRho(rho));
    float invN = rcp(lutSize);
    float3 uvw = gridCoord * ((lutSize-1.0f)*invN) + (0.5f*invN);
    float back = MaxEntResponseLUT.SampleLevel(
        LinearClampSampler, uvw, 0.0f).r;
    float eRaw = back + rho * max(m, 0.0f);
    return (omega / PI) * eRaw;
}
\end{lstlisting}

The entire shader performs a single texture sample---no quadrature or root-finding. R16F tables at $32^3$, $48^3$, and $64^3$ occupy $64$, $216$, and $512\,\text{KiB}$, respectively. Half-precision quantization errors are reported separately in the experiments chapter.


\section{ZH3/QZH as the Second-Order Truncation of Five-Parameter MaxEnt}

\subsection{What Should Be Truncated Is the Angular Energy Density}

The directional probability marginal is $f_{\text{dir}}(\mu)=g(\mu)^{-3}/(2\pi Q)$, but linear tensor moments and irradiance encoding correspond to the angular energy density weighted by the conditional mean energy:
\begin{equation*}
  \ell(\mu) = f_{\text{dir}}(\mu)\,\EE[r\mid\mu]
  = \frac{\omega}{2\pi W}\,g(\mu)^{-4},\quad \int_{S^2}\ell\dif\Omega = \omega,
\end{equation*}
where $Q=\int_{-1}^1 g(\mu)^{-3}\dif\mu$ and $W=\int_{-1}^1 g(\mu)^{-4}\dif\mu$ are both explicit closed forms. Defining normalized Legendre moments $\rho_l=\frac{1}{\omega}\int_{S^2}\ell(\mu)P_l(\mu)\dif\Omega$, the moment formulas of the previous section immediately give $\rho_0=1$, $\rho_1=v_\parallel/\omega=\rho$, $\rho_2=\zeta/\omega=\eta$.

\subsection{Second-Order Orthogonal Projection Theorem}

Legendre polynomials satisfy $\int_{-1}^1 P_l P_m\dif\mu = 2\delta_{lm}/(2l+1)$. Hence the orthogonal projection of $\ell$ onto the $l\leq 2$ space is strictly
\begin{equation*}
  \Pi_2[\ell](\mu) = \frac{\omega}{4\pi}\bigl(1 + 3\rho P_1(\mu) + 5\eta P_2(\mu)\bigr).
\end{equation*}
This is precisely a DC term, full linear SH, and a single axisymmetric quadratic coefficient along the axis $\mathbf{n}$: $1+3+1=5$ scalars per channel. Therefore, for this coaxial MaxEnt subfamily, ZH3/QZH is not a separate probability family to be asymptotically matched---it is the exact $l\leq 2$ orthogonal truncation of its angular energy density.

The high-frequency residual between the full density and the truncation is $\frac{\omega}{4\pi}\sum_{l=3}^\infty (2l+1)\rho_l P_l(\mu)$, where each $\rho_l$ can still be evaluated exactly from the elementary antiderivative of $g(\mu)^{-4}$ at the endpoints. What QZH discards are these high-frequency components that are nonlinearly constrained by $(\kappa,\alpha)$; the full MaxEnt cannot recover arbitrary unknown high frequencies, but it yields a specific closure that is consistent with the five stored moments and strictly positive.

\subsection{Canonical Correspondence with the Original ZH3 Paper}

Roughton~\etal's \textit{ZH3: Quadratic Zonal Harmonics}~\cite{Roughton2024ZH3} combines the four coefficients of linear SH with a single $l=2,m=0$ coefficient along a specified axisymmetric axis, yielding five coefficients per channel. Our $\mathbf{n}$ corresponds to that axisymmetric axis, and $\omega\rho_2$ corresponds to the axisymmetric second-order coefficient after basis-function normalization conversion. Distinguishing the three constructions in the paper: stored ZH3's axisymmetric coefficient is the true second-order projection of the signal (for our states, strictly equal to a normalized version of $\omega\rho_2$); curve-fit (hallucinated) ZH3 predicts the second-order coefficient empirically from the ratio of DC to linear SH (only equal to the true truncation when the prediction is exact); optimized or shared-luminance axis allows independent adjustment of the axis direction. Our five-parameter subfamily requires the dipole and second-order stress to be strictly coaxial.

\subsection{Truncation Positivity versus Full Positivity}

As long as $g_{\min}>0$, the full MaxEnt energy density $\ell\propto g^{-4}$ is strictly positive. The truncation polynomial $q(\mu)=1+3\rho_1\mu+5\rho_2 P_2(\mu)$, however, is not guaranteed to be non-negative. Taking the axis orientation such that $\rho_1\geq 0$, its necessary and sufficient non-negativity condition is
\begin{equation*}
  \begin{cases}
    1-3\rho_1+5\rho_2\geq 0 & \rho_2\leq 0\ \text{or}\ \rho_1\geq 5\rho_2,\\[4pt]
    1-\dfrac{5\rho_2}{2}-\dfrac{3\rho_1^2}{10\rho_2}\geq 0 & \rho_2>0\ \text{and}\ 0\leq\rho_1\leq 5\rho_2.
  \end{cases}
\end{equation*}
This is the same quadratic polynomial positivity problem as Appendix~B Eqs.~(22)--(26) of the original paper, after normalization conversion. The paper's bounds constrain the truncated ZH3 reconstruction itself; our $g_{\min}>0$ constraint bounds the untruncated MaxEnt density. When the former fails it produces negative lobes and ringing; the latter remains positive.


\section{Experimental Comparison: Irradiance Reconstruction Accuracy of MaxEnt vs.\ SH/QZH}

This chapter first conducts a large-scale normalized-cosine irradiance benchmark on 981 Poly Haven HDRI~2K scenes, followed by small-sample controls on three Debevec probes and four analytic axisymmetric spotlight parameters. Each experimental group uses a consistent reference, error metric, and implementation; different groups respectively adopt compressed directional atoms, Fibonacci spherical integration, and analytic high-order quadrature as references.

\subsection{Experimental Setup}

\subsubsection{Methods Compared}

Six methods, each reconstructing RGB channels independently using $E(\mathbf{n})=\pi^{-1}\int L(\mathbf{u})\max(\mathbf{n}\cdot\mathbf{u},0)\dif\Omega$:
\begin{itemize}
  \item \textbf{SH-1 (ZH3 paper's SH2)}: DC and $l=1$, 4 DOF per channel;~\cite{Ramamoorthi2001Efficient,Sloan2008StupidSH}
  \item \textbf{SH-2 (ZH3 paper's SH3)}: full $l\leq 2$, 9 DOF per channel;
  \item \textbf{Stored QZH}: stores the true $l=2$ axisymmetric projection along each channel's own first-moment axis, 5 DOF per channel;~\cite{Roughton2024ZH3}
  \item \textbf{Curve-fit QZH}: predicts the axisymmetric coefficient from the DC and $l=1$ ratio via a curve-fit formula, 4 DOF per channel;~\cite{Roughton2024ZH3}
  \item \textbf{MaxEnt-4}: pure dipole model, 4 DOF per channel;
  \item \textbf{MaxEnt-5}: dipole and second-order stress coaxial subfamily, 5 DOF per channel.
\end{itemize}

The current implementation computes axes independently for each RGB channel and does not employ the shared-luminance axis discussed in the ZH3 literature. Grouped by DOF: SH-1, curve-fit QZH, and MaxEnt-4 are 4 DOF; stored QZH and MaxEnt-5 are 5 DOF; SH-2 is 9 DOF.

\subsubsection{Reference Ground Truth and Error Metric}

HDR probe references use $N=24576$ deterministic Fibonacci spherical directions~\cite{Marques2013Fibonacci} for quasi-uniform spherical quadrature. Analytic spotlights are computed by high-order Gauss--Legendre quadrature. Per-channel error is normalized by the channel's total energy proxy $\omega_c=2\sqrt{\pi}\,\text{SH}[0]_c$:
\begin{equation*}
  \text{RMSE}_{\text{norm}}=\sqrt{\frac{1}{M}\sum_{i=1}^M\Bigl(\frac{E_{\text{method}}(\mathbf{n}_i)-E_{\text{ref}}(\mathbf{n}_i)}{\omega_c}\Bigr)^2}.
\end{equation*}
Both HDR and spotlight evaluations use 4096 Fibonacci spherical directions for the normal. CSVs record \texttt{negative\_fraction} for each R/G/B channel separately; the ``RGB average negative fraction'' in figures is the arithmetic mean of the three per-channel negative fractions.

\subsection{Large-Scale Benchmark: Poly Haven HDRI~2K}

981 equirectangular HDRs ($2048\times 1024$ EXR, covering outdoor, indoor, night, sun, and point-light scenes). The reference uses energy-conserving pixel-block compression: each HDR is first partitioned into $8\times 8$ pixel blocks; blocks whose peak luminance exceeds 8$\times$ the block mean are recursively split into $2\times 2$ sub-blocks, yielding approximately 33k directional atoms (about 1.6\% of the 2.1M pixels). Evaluation normals are 1024 Fibonacci directions.

\subsubsection{Overall Results}

\begin{figure}[htb]
  \centering
  \includegraphics[width=\columnwidth]{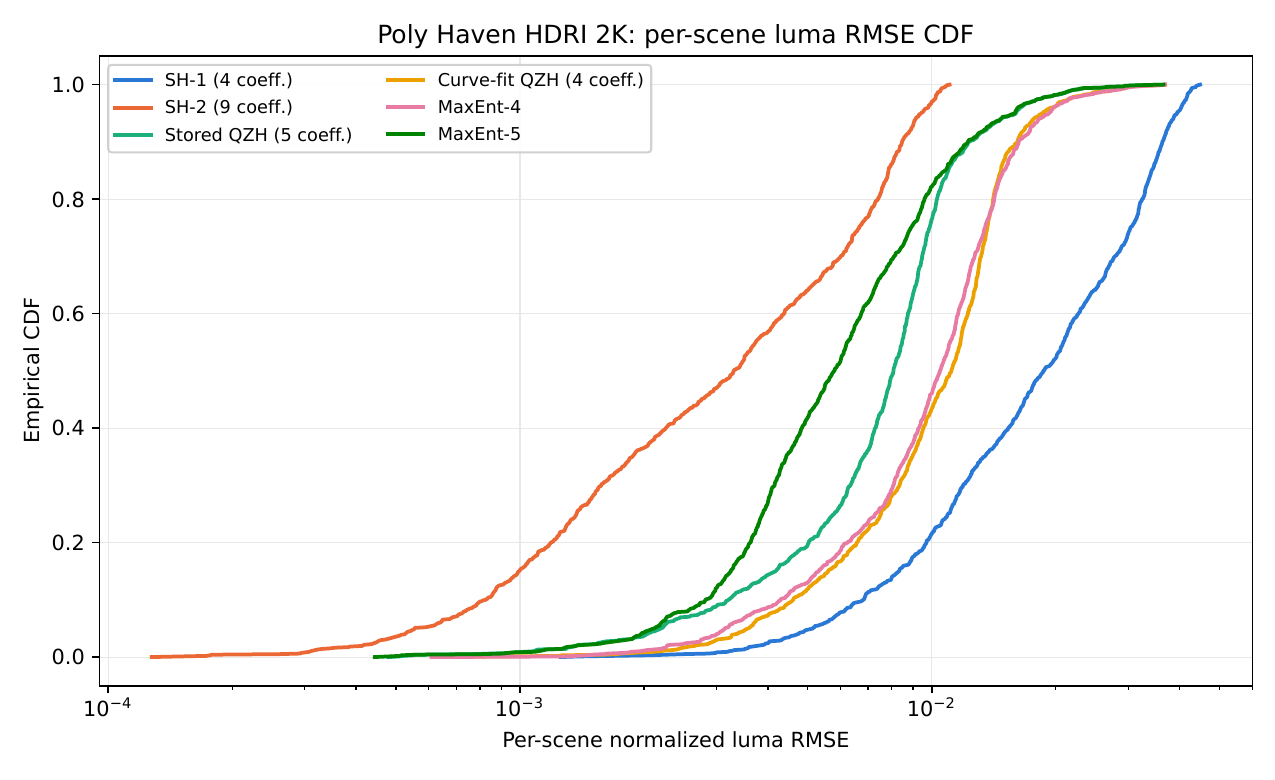}
  \caption{Empirical CDF of per-scene luminance RMSE across 981 Poly Haven HDRIs (logarithmic horizontal axis). The distributions of SH-2 and MaxEnt-5 lie below the other methods overall.}
  \label{fig:ph-cdf}
\end{figure}

\begin{figure*}[tb]
  \centering
  \includegraphics[width=\textwidth]{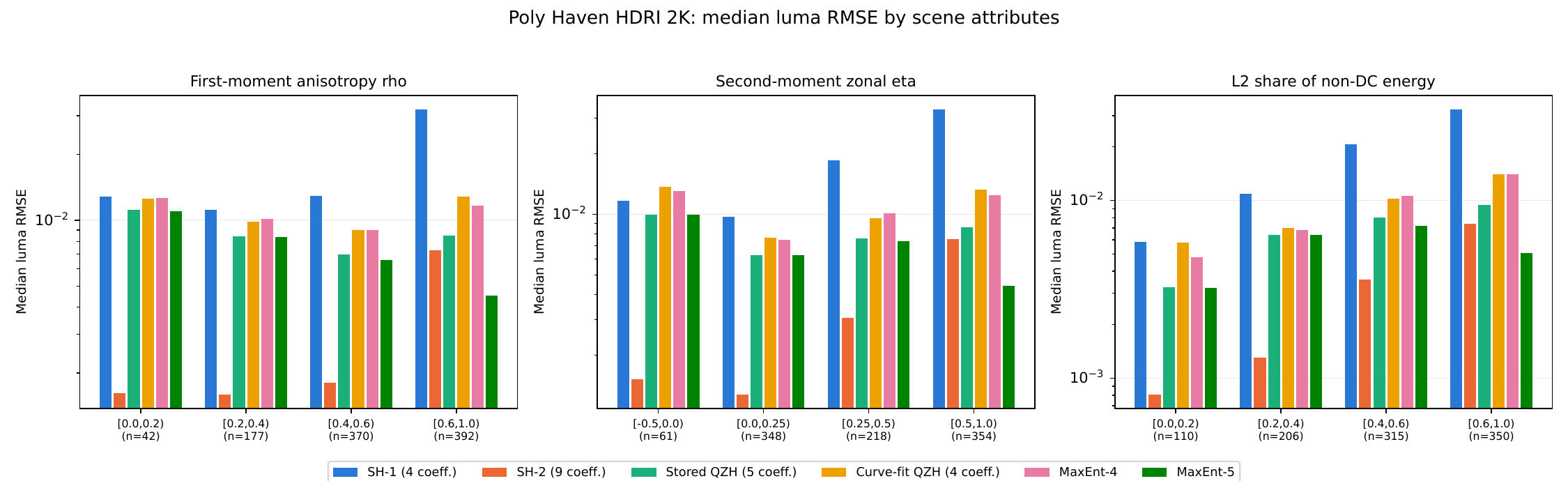}
  \caption{Median luminance RMSE bucketed by scene property. Left: first-order anisotropy $\rho$; Middle: axisymmetric second moment $\eta$; Right: second-band fraction of non-DC spherical harmonic energy.}
  \label{fig:ph-buckets}
\end{figure*}

\begin{table*}[tb]
  \centering
  \caption{Per-scene luminance RMSE percentiles and mean across 981 Poly Haven HDRIs}
  \label{tbl:ph-summary}
  \begin{tabular}{lcccccc}
    \toprule
    Method & Median & p75 & p90 & p95 & p99 & Mean \\
    \midrule
    SH-1 & 0.0186 & 0.0304 & 0.0363 & 0.0394 & 0.0425 & 0.0205 \\
    SH-2 & 0.0033 & 0.0067 & 0.0085 & 0.0094 & 0.0105 & 0.0041 \\
    Stored QZH & 0.0081 & 0.0099 & 0.0123 & 0.0160 & 0.0219 & 0.0082 \\
    Curve-fit QZH & 0.0112 & 0.0136 & 0.0161 & 0.0186 & 0.0273 & 0.0109 \\
    MaxEnt-4 & 0.0105 & 0.0135 & 0.0163 & 0.0193 & 0.0276 & 0.0107 \\
    MaxEnt-5 & 0.0058 & 0.0089 & 0.0122 & 0.0159 & 0.0219 & 0.0069 \\
    \bottomrule
  \end{tabular}
\end{table*}

SH-2 (9 DOF) achieves the best overall error on this dataset; MaxEnt-5 ranks second, with both median and mean approximately $0.7\times$ those of stored QZH. SH-1 has the highest p99 (0.0425) among all methods.

\subsubsection{Paired Comparisons}

Per-scene luminance RMSE differences $\Delta_i$ (negative = MaxEnt better) for MaxEnt-5 vs.\ stored QZH: MaxEnt-5 wins on 772 of 981 scenes (78.7\%), mean $\Delta=-1.29\times 10^{-3}$ (2000-sample bootstrap 95\% CI $[-1.42,-1.17]\times 10^{-3}$, excludes zero); Wilcoxon signed-rank $p\approx 5\times 10^{-96}$, sign test $p\approx 2\times 10^{-76}$. MaxEnt-4 vs.\ curve-fit QZH: wins on 513 scenes (Wilcoxon $p=0.035$); the two are roughly on par for natural lighting.

\begin{figure*}[tb]
  \centering
  \includegraphics[width=\textwidth]{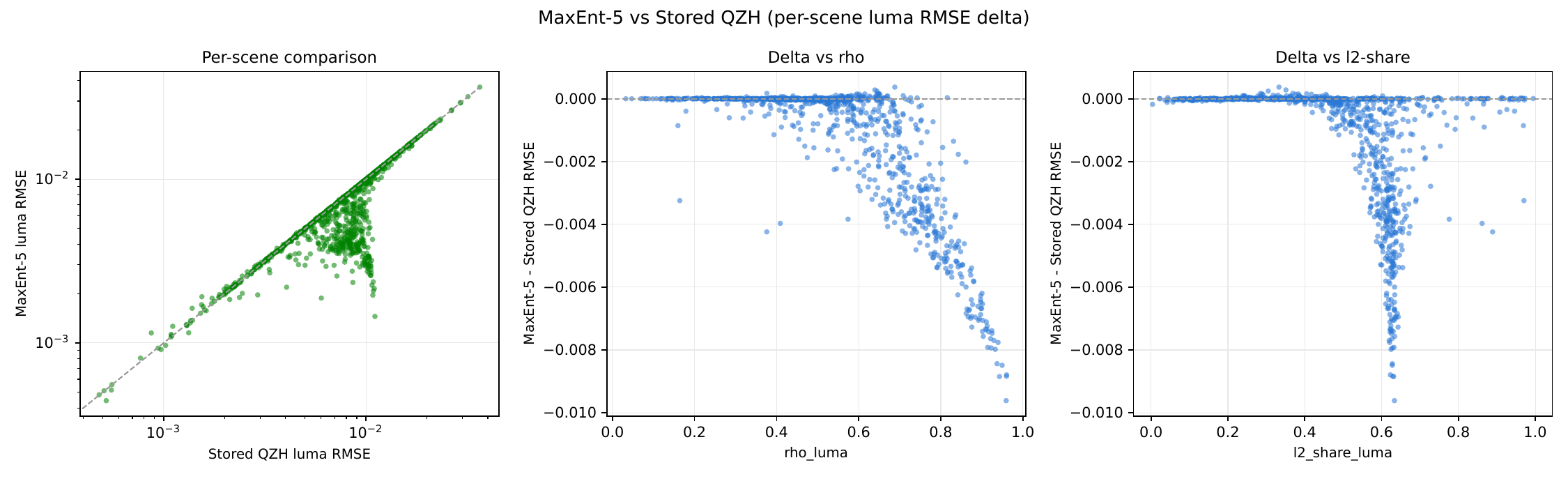}
  \caption{Per-scene comparison of MaxEnt-5 vs.\ stored QZH. Left: RMSE scatter of the two methods (dashed line = equality); Middle, Right: difference vs.\ $\rho$ and second-band energy fraction.}
  \label{fig:ph-delta}
\end{figure*}

\subsubsection{Error by Scene Property}

Bucket statistics show that the MaxEnt advantage over QZH increases monotonically with directionality (Fig.~\ref{fig:ph-delta}): win rate 42.9\% for $\rho\in[0,0.2)$, 91.3\% for $\rho\in[0.6,1)$ (392 scenes), with mean $\Delta$ changing from $-1.1\times 10^{-4}$ to $-3.0\times 10^{-3}$; win rate 88.9\% for the bucket with second-band fraction of non-DC energy in $[0.6,1)$. In the strong-directionality bucket MaxEnt-5's median RMSE is 0.0045, approximately $0.53\times$ that of stored QZH (0.0085), and lower than full SH-2 (0.0073). Paired median $\Delta=-2.9\times 10^{-3}$ in this bucket. In the $\eta\in[0.5,1)$ bucket (354 scenes), MaxEnt-5 median 0.0044, a 48.7\% reduction relative to QZH (0.0086) and a 41.4\% reduction relative to SH-2 (0.0076).

\subsubsection{Negative Values and Worst-Case Scenes}

MaxEnt-4 and MaxEnt-5 exhibit zero negative irradiance across all 981 scenes; SH-1 has negative values on 579 scenes (mean 7.1\%, max 24\%), SH-2 on 120 scenes (max 25\%), and the two QZH variants on approximately 52--60 scenes each (max 25\%--28\%). Bucketed by $\rho$, the negative fraction of SH-1 rises with directionality; all 392 scenes in the $\rho\in[0.6,1)$ bucket exhibit negatives, while both MaxEnt variants remain zero across all buckets.

The 20 worst scenes ranked by MaxEnt-5 luminance RMSE are predominantly tunnels and indoor scenes (e.g., concrete\_tunnel\_02, short\_tunnel, studio\_small\_01). These scenes contain multiple non-coaxial reflected light sources; the axisymmetric assumptions of both the five-parameter coaxial model and QZH cannot express their second-order structure, and the two have similar errors (approximately 0.0197--0.0365) that are far higher than SH-2 (approximately 0.0027--0.0086)---a systematic limitation of the coaxial subfamily.

\subsubsection{GPU Decoding Micro-Benchmark}

RTX 4060 Laptop GPU (driver 32.0.16.1062, high-performance power plan), Rust + wgpu~24 + Direct3D~12 backend, no window, no swap chain, no vertical sync; GPU-only time measured via timestamp queries. Parameter textures, output textures, and LUTs are created once; each configuration is pre-warmed with $\geq$1000 dispatches and $\geq$2 seconds, followed by 3 repeated experiments of 100 rounds each. Three parameter modes: random (independent random per pixel, cache-pressure upper bound), coherent (low-frequency smooth), and boundary ($\rho\to 1$, $q\to 0|1$, $m$ crosses zero). LUTs use R16F 3D textures with trilinear sampling, tested at 32, 48, and 64 resolutions.

\begin{table*}[tb]
  \centering
  \caption{GPU-only time in $\mu\mathrm{s}$ per dispatch. Each sample is the mean of a dispatch batch; the table reports the median across samples. QZH column = absolute median; three $\Delta$ columns = paired medians relative to stored QZH. Host encoding and submission cost approximately $175\,\mu\mathrm{s}$/round.}
  \label{tbl:gpu-decoder}
  \small
  \begin{tabular}{llrrrr}
    \toprule
    Resolution & Mode & QZH median & $\Delta\ 32^3$ & $\Delta\ 48^3$ & $\Delta\ 64^3$ \\
    \midrule
    $1920\times1080$ & Random & 279.3 & $+0.9$ & $-1.8$ & $+164.1$ \\
    $1920\times1080$ & Coherent & 127.8 & $-8.2$ & $-8.4$ & $-8.1$ \\
    $1920\times1080$ & Boundary & 151.2 & $+6.3$ & $+9.4$ & $+9.9$ \\
    $2560\times1600$ & Random & 562.9 & $+1.2$ & $-9.1$ & $+331.3$ \\
    $2560\times1600$ & Coherent & 229.8 & $+10.9$ & $+12.2$ & $+14.4$ \\
    $2560\times1600$ & Boundary & 296.1 & $+30.3$ & $+34.5$ & $+36.7$ \\
    $3840\times2160$ & Random & 1121.3 & $-0.3$ & $+15.2$ & $+687.6$ \\
    $3840\times2160$ & Coherent & 443.6 & $+32.7$ & $+36.7$ & $+45.6$ \\
    $3840\times2160$ & Boundary & 610.1 & $+71.8$ & $+82.2$ & $+87.1$ \\
    \bottomrule
  \end{tabular}
\end{table*}

In random mode, each 4K dispatch reads three RGBA16F parameter textures totaling $\sim$200 MB; absolute times are notably higher than coherent mode. QZH and MaxEnt share this input cost, so paired deltas more directly reflect decoder differences. Paired bootstrap 95\% intervals use 300 paired samples; all 27 combinations have interval width below $10\,\mu\mathrm{s}$. Three observations: (1) in coherent and boundary modes, the three LUT deltas are close; (2) in random mode, $32^3$ deltas are near zero, $48^3$ range from $-9.1$ to $+15.2\,\mu\mathrm{s}$, and $64^3$ grows with resolution to $+687.6\,\mu\mathrm{s}$---consistent with random trilinear access crossing cache hierarchy levels; (3) in coherent mode the three LUT deltas at 4K are $32.7$--$45.6\,\mu\mathrm{s}$. This benchmark is an isolated decoder cost; it does not include probe interpolation or a full frame pipeline.

\subsection{LUT Accuracy Verification}

\subsubsection{Exact Reference and Convergence}

The exact back response $B_{\text{exact}}(\rho,\eta,u)=e_{\text{raw}}(-u)$ is computed from the azimuthal algebraic closed form of the angular density $g(\mu)^{-4}$ plus an outer one-dimensional Gauss--Legendre integral. The inverse mapping from moment parameters to natural parameters uses vectorized damped Newton (99.8\% of uniform-domain points converge). Two integration node budgets (256 and 1024) differ by less than $10^{-12}$ on 99.7\% of test points. Near-boundary ill-conditioned points ($g_{\min}<10^{-3}$, approximately 0.3\%) are filtered out and excluded from statistics.

\subsubsection{Domain-Sampled Error Matrix}

\begin{table*}[tb]
  \centering
  \caption{LUT back-response absolute error (R32F) on three sampling domains. Error decreases with resolution; maxima occur in the boundary degenerate region.}
  \label{tbl:lut-error}
  \small
  \begin{tabular}{llrrrrr}
    \toprule
    Domain & LUT & RMSE & p95 & p99 & p99.9 & Max \\
    \midrule
    Physically uniform & $32^3$ & 5.5e-3 & 1.3e-2 & 2.2e-2 & 5.2e-2 & 7.8e-2 \\
    Physically uniform & $48^3$ & 4.2e-3 & 8.9e-3 & 1.7e-2 & 4.4e-2 & 7.0e-2 \\
    Physically uniform & $64^3$ & 3.4e-3 & 6.8e-3 & 1.4e-2 & 3.9e-2 & 6.4e-2 \\
    Texture uniform & $32^3$ & 6.9e-3 & 1.6e-2 & 3.1e-2 & 4.9e-2 & 1.7e-1 \\
    Texture uniform & $48^3$ & 5.6e-3 & 1.3e-2 & 2.5e-2 & 4.3e-2 & 1.7e-1 \\
    Texture uniform & $64^3$ & 4.8e-3 & 1.1e-2 & 2.1e-2 & 3.9e-2 & 1.7e-1 \\
    Boundary layer & $32^3$ & 1.8e-2 & 2.5e-2 & 6.7e-2 & 1.9e-1 & 2.2e-1 \\
    Boundary layer & $48^3$ & 1.6e-2 & 2.1e-2 & 5.8e-2 & 1.9e-1 & 2.2e-1 \\
    Boundary layer & $64^3$ & 1.6e-2 & 1.8e-2 & 5.3e-2 & 1.9e-1 & 2.2e-1 \\
    \bottomrule
  \end{tabular}
\end{table*}

\begin{figure*}[b]
  \centering
  \includegraphics[width=0.58\textwidth]{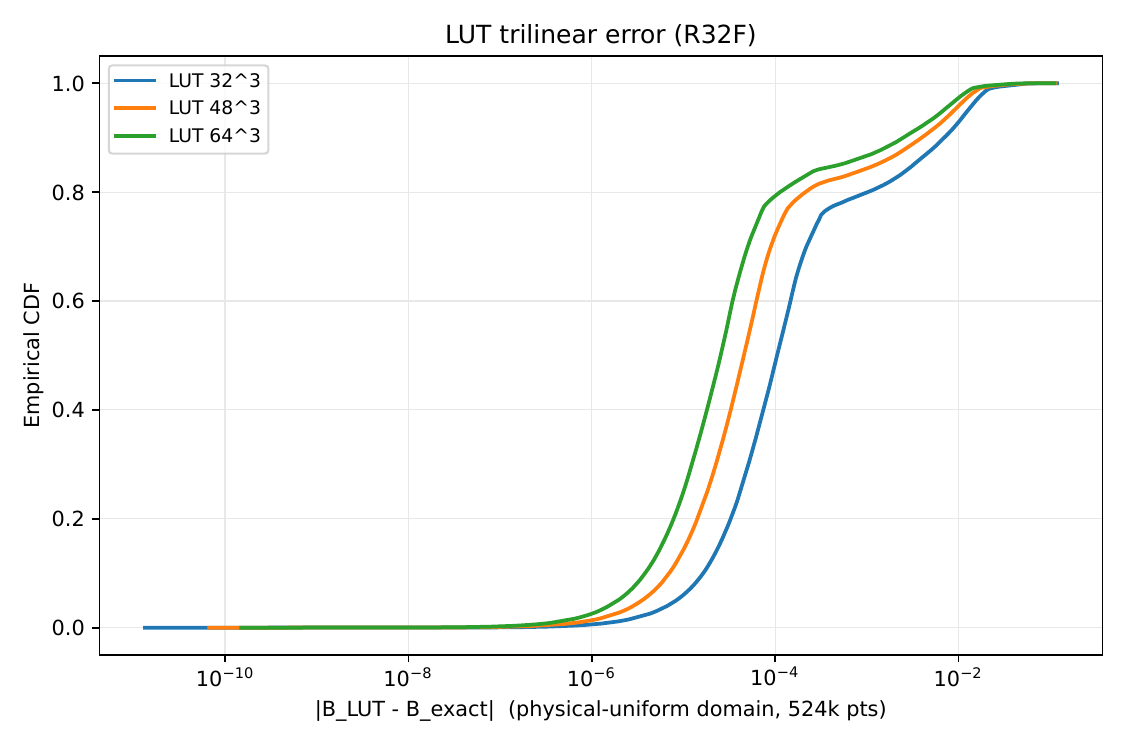}
  \captionsetup{width=0.78\textwidth}
  \caption{Empirical CDF of LUT error on the physically uniform domain.}
  \label{fig:lut-cdf}
\end{figure*}

\begin{figure*}[tb]
  \centering
  \includegraphics[width=\textwidth]{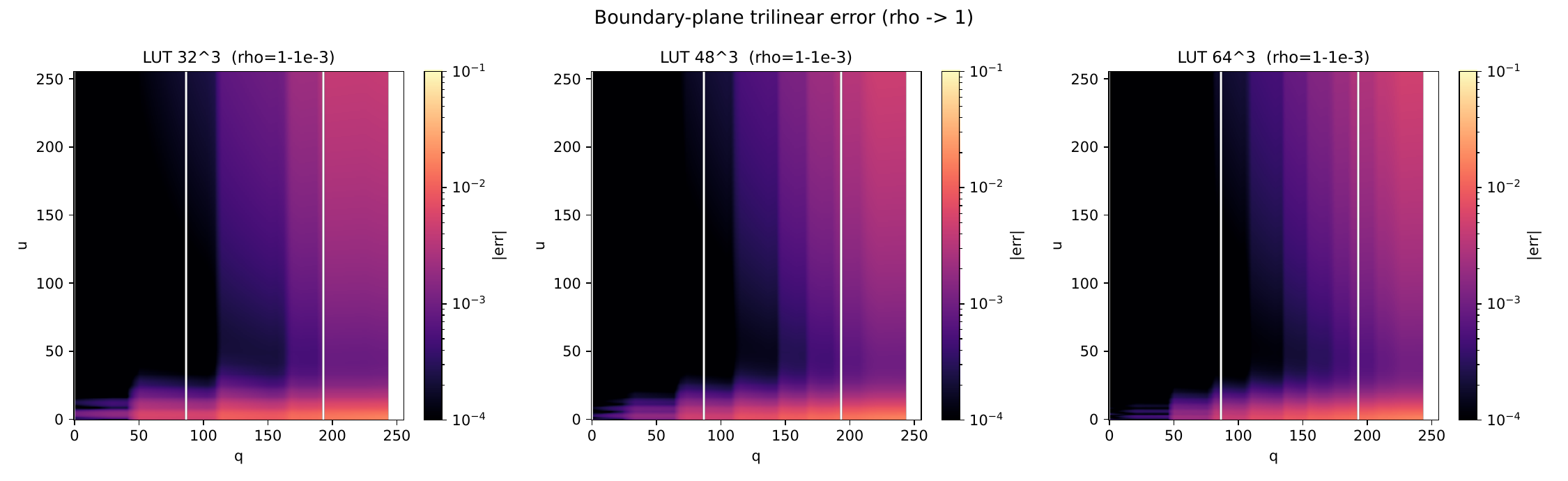}
  \caption{LUT error on the $q\times u$ grid (in sample-index coordinates) at the $\rho=1-10^{-3}$ surface (logarithmic color scale). Error concentrates in the $q\to 1$ degenerate region.}
  \label{fig:lut-heatmap}
\end{figure*}

Maximum deviation at the isotropic point: $6.8\times 10^{-5}$ ($32^3$); maximum error of the pure-dipole-subfamily LUT relative to the analytic cosine formula: $3.7\times 10^{-3}$ ($32^3$); antisymmetry identity $e(m)-e(-m)=\rho m$ residual: $1.1\times 10^{-16}$. GPU/CPU consistency: mean difference $2.9\times 10^{-5}$, 99th percentile $2.0\times 10^{-4}$.

\subsubsection{981-Scene End-to-End and Benefit Retention}

The end-to-end comparison shares the same 1024 Fibonacci normals and the same compressed-atom reference as the main Poly Haven benchmark; QZH, exact MaxEnt, and three R16F LUTs are all recomputed on this grid.

\begin{table*}[tb]
  \centering
  \caption{End-to-end luminance RMSE and win rate relative to QZH across 981 Poly Haven scenes}
  \label{tbl:lut-e2e}
  \begin{tabular}{lrrr}
    \toprule
    Method & Mean RMSE & Median & Win rate vs.\ QZH \\
    \midrule
    Stored QZH & 0.00819 & 0.00807 & -- \\
    Exact MaxEnt & 0.00690 & 0.00580 & 78.7\% \\
    LUT $32^3$ & 0.00690 & 0.00580 & 75.8\% \\
    LUT $48^3$ & 0.00690 & 0.00580 & 77.2\% \\
    LUT $64^3$ & 0.00690 & 0.00580 & 78.5\% \\
    \bottomrule
  \end{tabular}
\end{table*}

\begin{figure*}[tb]
  \centering
  \includegraphics[width=0.62\textwidth]{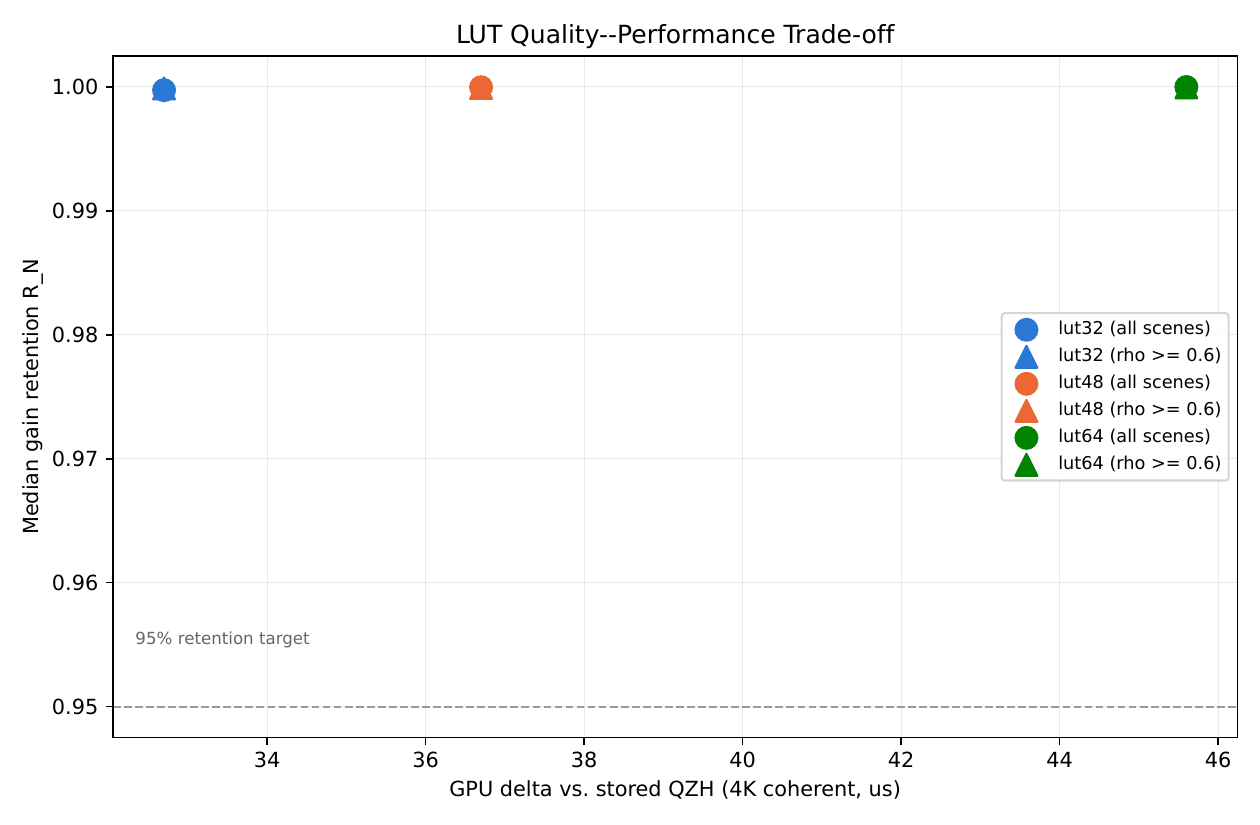}
  \captionsetup{width=0.78\textwidth}
  \caption{Quality--performance Pareto: horizontal axis = GPU delta relative to QZH in 4K coherent mode, vertical axis = median benefit retention rate. Circles = all scenes, triangles = $\rho\geq 0.6$ bucket. All three resolutions meet the 95\% retention target.}
  \label{fig:lut-pareto}
\end{figure*}

Per-scene RMSE differences between LUT and exact MaxEnt are at the $10^{-6}$ scale (5th decimal place). Median benefit retention rates are 0.9997 ($32^3$) and 1.0000 ($48^3$/$64^3$); in the critical $\rho\geq 0.6$ advantage bucket all medians are at least 0.9998, with $R<0.9$ scene fractions of 0.84\%, 0.28\%, and 0\%, respectively. Considering exact reference convergence, interpolation error relative to model benefit, median and high-directionality-bucket benefit retention, non-negativity and finite-value checks, p99 error, and GPU cost, $32^3$ already satisfies our production criterion; higher resolutions yield negligible end-to-end accuracy gains.
\FloatBarrier

\subsection{HDR Light Probe Benchmark}

Three standard Debevec angular-map HDR environment light probes~\cite{Debevec1998Rendering,DebevecLightProbes}: Grace Cathedral, St.\ Peter's Basilica, Uffizi Gallery.

\begin{figure*}[tb]
  \centering
  \includegraphics[width=\textwidth]{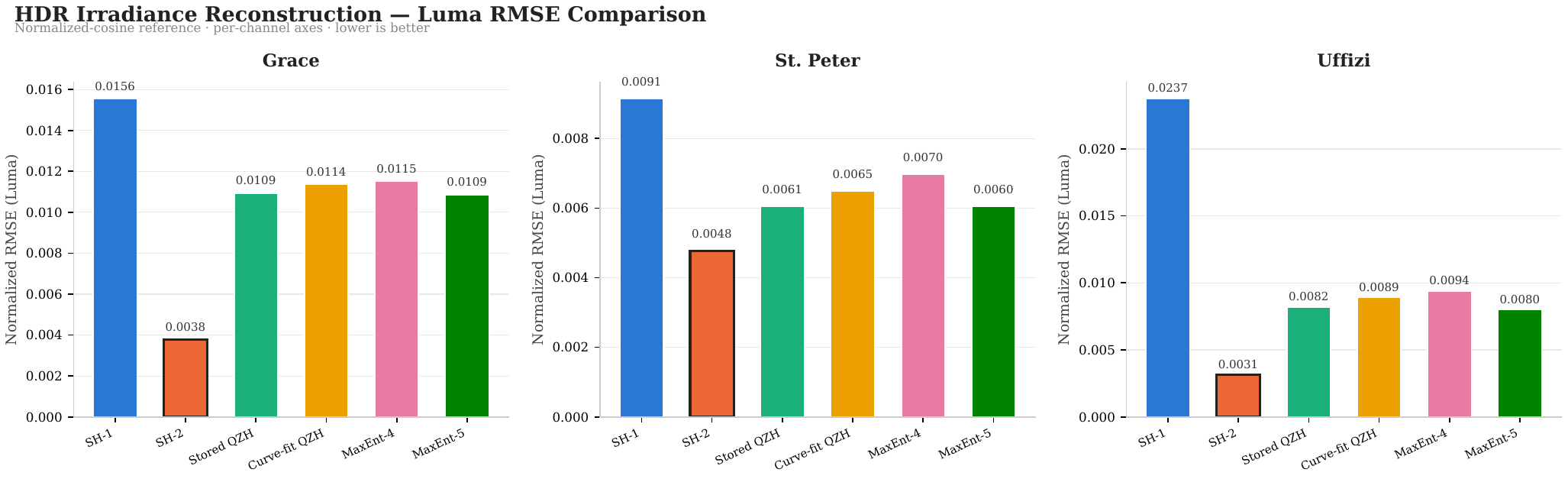}
  \caption{Luminance RMSE of the six methods on the three HDR probes. SH-2 is lowest on this dataset; five-parameter MaxEnt is close to stored QZH.}
  \label{fig:hdr-bars}
\end{figure*}

\begin{figure*}[tb]
  \centering
  \includegraphics[width=\textwidth]{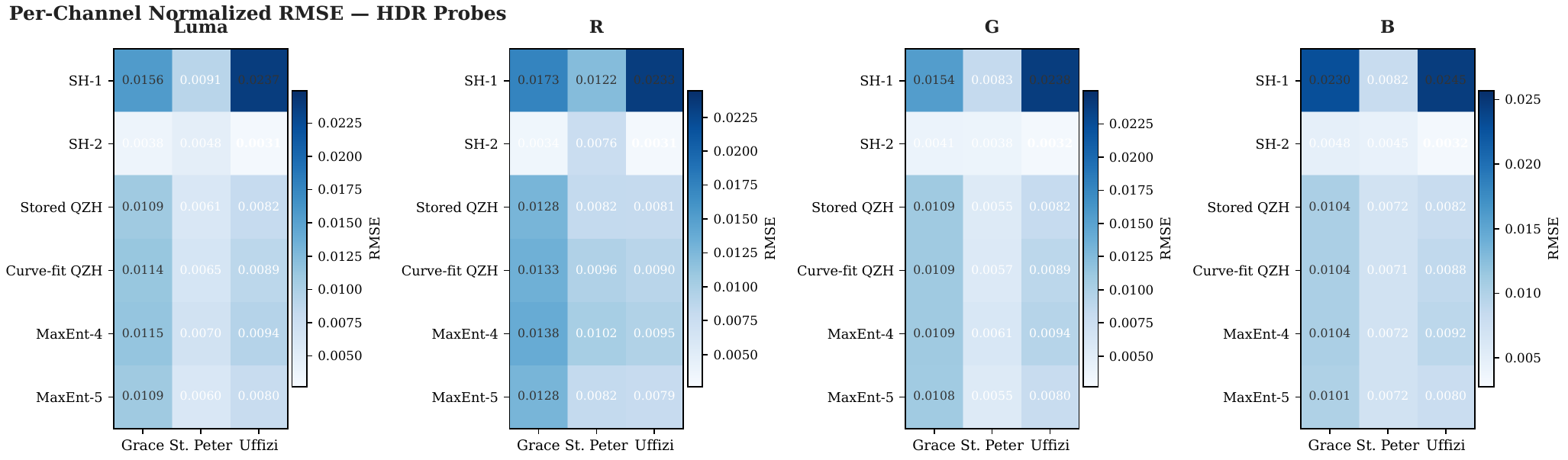}
  \caption{Per-channel RMSE heatmap for the HDR probes. SH-2 is consistently best across all channels; five-parameter MaxEnt is close to stored QZH.}
  \label{fig:hdr-heatmap}
\end{figure*}

Key luminance RMSE findings: SH-2 achieves the lowest RMSE on all three HDR probes (using the full second-order space, at the cost of higher storage and possible negative channels); among 5-DOF methods, five-parameter MaxEnt and stored QZH are numerically close (Grace 0.0109/0.0109, St.\ Peter 0.0060/0.0061, Uffizi 0.0080/0.0082); among 4-DOF methods, curve-fit QZH is slightly lower than MaxEnt-4, and both are lower than SH-1 (MaxEnt-4's RMSE reduction relative to SH-1 is approximately 24\%--61\%).

\begin{figure*}[b]
  \centering
  \includegraphics[width=0.82\textwidth]{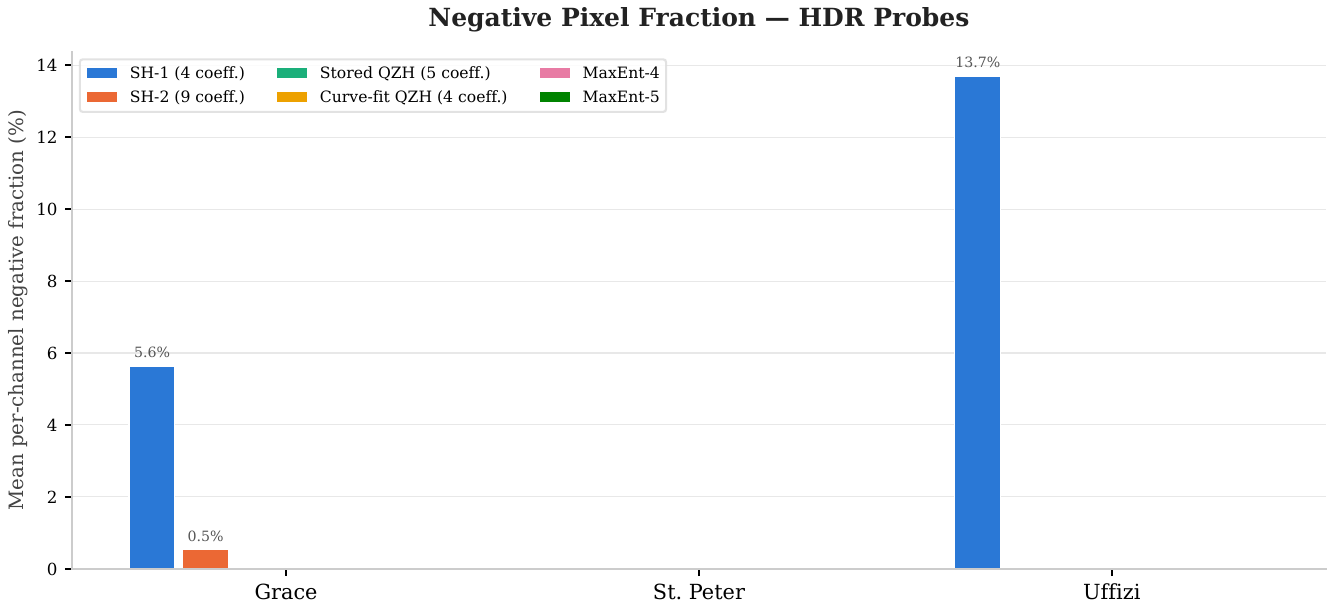}
  \caption{Average RGB negative-channel pixel fraction for each method on the HDR probes. SH-1 produces substantial negatives on Grace (5.6\%) and Uffizi (13.7\%); SH-2 has a small fraction only on Grace (0.5\%); QZH and both MaxEnt variants are 0\% on all tests.}
  \label{fig:neg-frac}
\end{figure*}

The target quantity irradiance is non-negative. SH-1 exhibits negative channels on probes with strong anisotropic components, and SH-2 also lacks an unconditional positivity guarantee. The full MaxEnt density has an analytic positivity guarantee when $g_{\min}>0$; QZH and curve-fit QZH are finite-order polynomials that only guarantee non-negativity when their respective quadratic positivity conditions are satisfied. This experiment does not apply any generic positivity fix to QZH; the observed zero-negative fraction is a data-dependent outcome.
\FloatBarrier

\subsection{Axisymmetric Spotlight Sweep}

Analytic axisymmetric spherical-Gaussian spotlights $\ell(\mu)\propto\exp(\lambda(\mu-1))$, $\lambda\in\{16,32,64,128\}$. In this setting: the true distribution and the coaxial MaxEnt share a common symmetry axis, but the spherical Gaussian $\exp(\lambda(\mu-1))$ and MaxEnt's $g(\mu)^{-4}$ are not the same function family---the five-parameter result is purely the closure error after matching the first two moments; SH-2 and stored QZH (which stores the true second-order projection along the same axis) are analytically equivalent in this axisymmetric setting.

\begin{figure*}[tb]
  \centering
  \includegraphics[width=\textwidth]{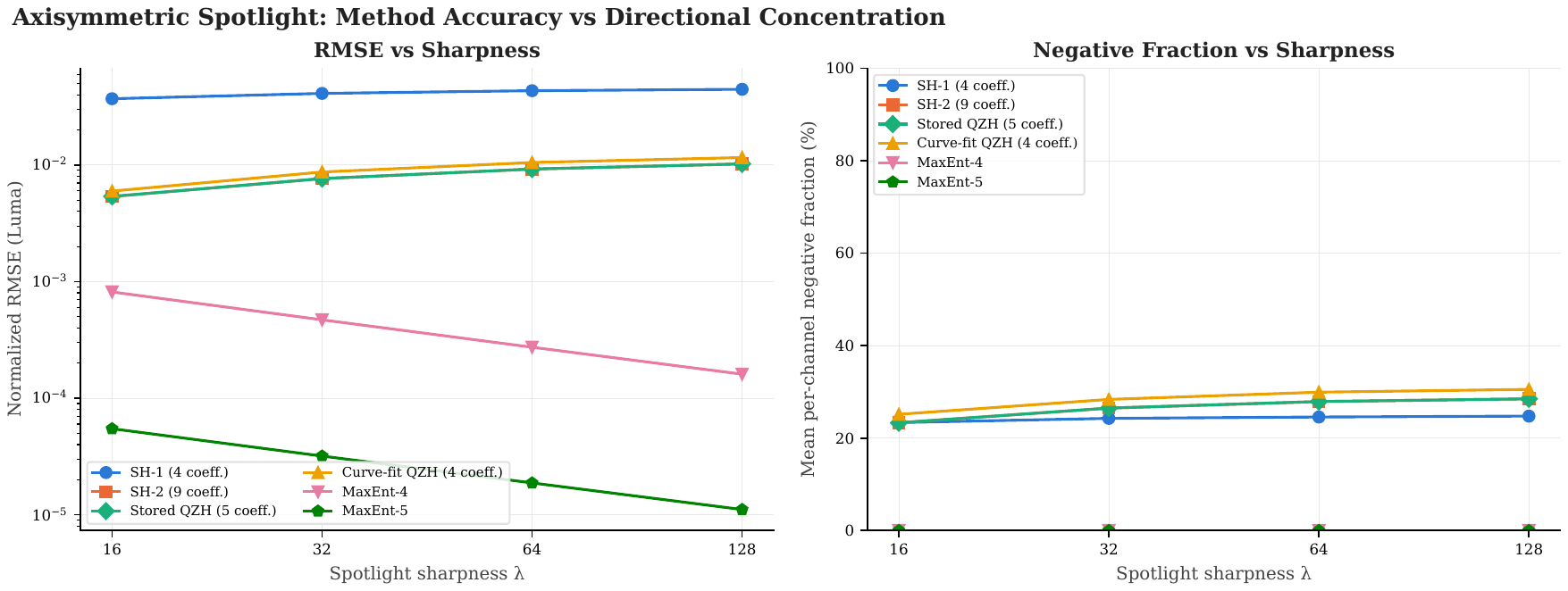}
  \caption{Axisymmetric spotlight RMSE and negative pixel fraction as a function of the directionality parameter $\lambda$. Left: luminance RMSE (logarithmic vertical axis). Right: negative pixel fraction. The MaxEnt family has zero negative pixels at all four sharpness levels.}
  \label{fig:spotlight}
\end{figure*}

\begin{strip}
  \centering
  \captionsetup{hypcap=false}
  \captionof{table}{Axisymmetric spotlight luminance RMSE summary. All MaxEnt variants have zero negative pixel fraction, while QZH reaches 28.5\% at $\lambda=128$.}
  \label{tab:spotlight_rmse}
  \small
  \begin{tabular}{lcccccc}
    \toprule
    $\lambda$ & SH-1 & SH-2/Stored QZH & Curve-fit QZH & MaxEnt-4 & MaxEnt-5 & MaxEnt-5 vs.\ QZH \\
    \midrule
    16 & 0.03706 & 0.00538 & 0.00598 & 0.00082 & \textbf{0.000055} & $98\times$ better \\
    32 & 0.04116 & 0.00765 & 0.00872 & 0.00047 & \textbf{0.000032} & $240\times$ better \\
    64 & 0.04343 & 0.00924 & 0.01052 & 0.00027 & \textbf{0.000019} & $492\times$ better \\
    128 & 0.04464 & 0.01023 & 0.01159 & 0.00016 & \textbf{0.000011} & $922\times$ better \\
    \bottomrule
  \end{tabular}
\end{strip}

Key observations: (1) At $\lambda=128$, five-parameter MaxEnt's luminance RMSE is approximately $1.1\times 10^{-5}$, stored QZH's is approximately $1.0\times 10^{-2}$, a factor of approximately 922$\times$ difference from unrounded data; four-parameter MaxEnt at approximately $1.6\times 10^{-4}$ is also notably lower than QZH; (2) finite SH/QZH truncations exhibit significant negatives at $\lambda=128$ (approximately 25\%--31\%), while both MaxEnt variants remain non-negative.

\subsection{Error Distribution}

\begin{figure*}[tb]
  \centering
  \includegraphics[width=\textwidth]{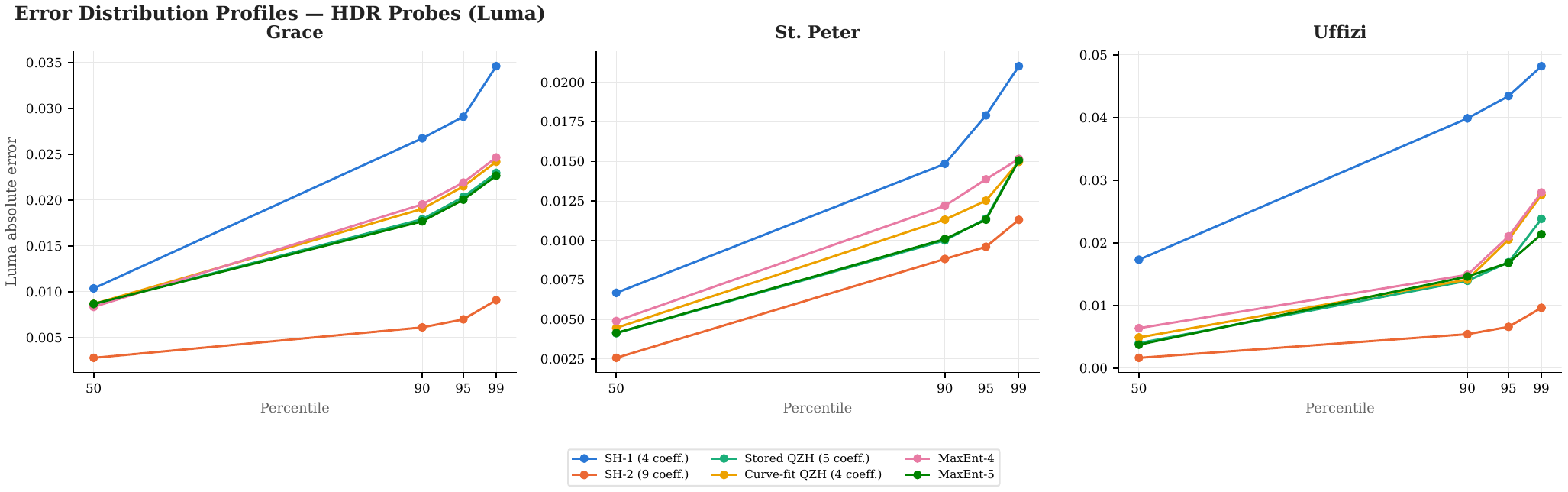}
  \caption{Luminance absolute error percentile curves for the three HDR probes. Horizontal axis = percentile (50th, 90th, 95th, 99th); vertical axis = corresponding normalized absolute error.}
  \label{fig:err-dist}
\end{figure*}

SH-2 has lower p50 and p99 than all other methods on all three probes; SH-1 exhibits a thicker error tail on this dataset (Uffizi's p99 is approximately $2.78\times$ its p50); several error percentiles of stored QZH and five-parameter MaxEnt are close, consistent with their shared low-order axial moment construction.
\clearpage

\subsection{Visual Comparison}

Fig.~\ref{fig:vis} collects the reconstructions of the three HDR probes under each method, marking pixels with negative channels in pink, thereby visually exposing spatial structure and positivity differences beyond numerical error.

\begin{strip}
  \centering
  \includegraphics[width=\textwidth]{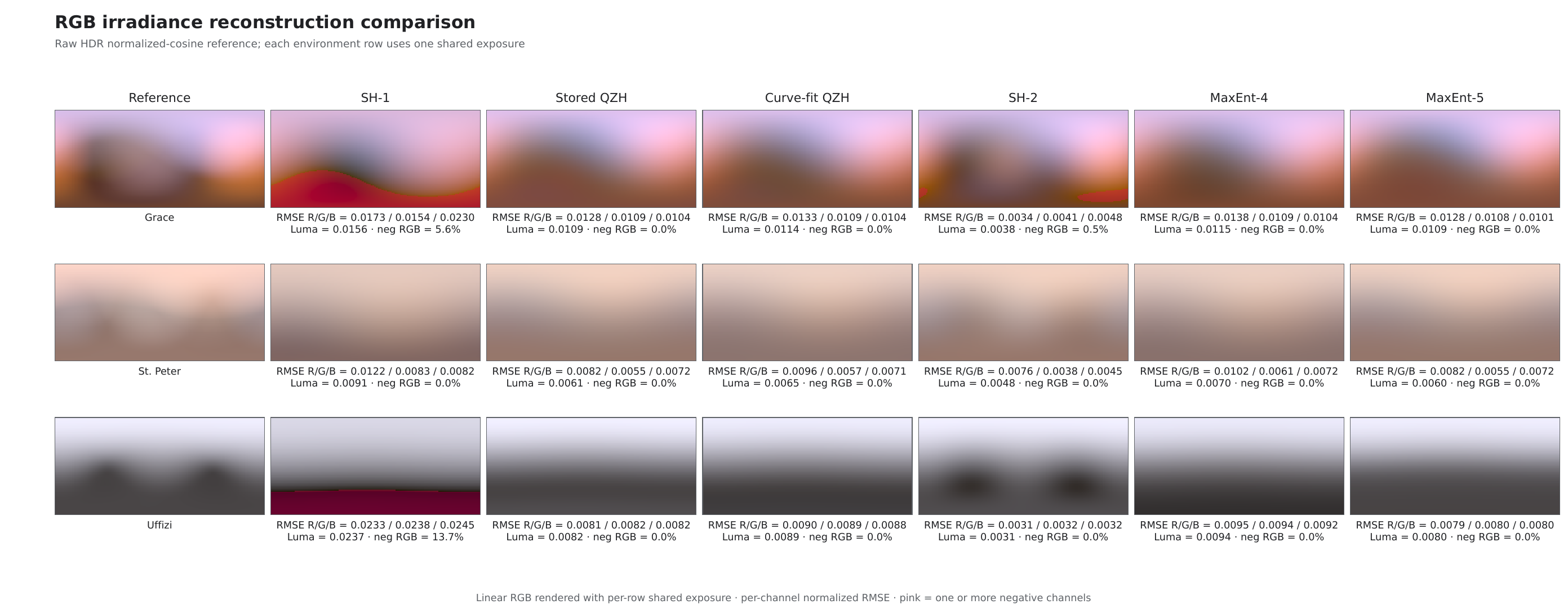}
  \captionsetup{hypcap=false}
  \captionof{figure}{Visual comparison of RGB irradiance reconstruction on the HDR probes. Each row is one probe, each column one method. Pink pixels indicate at least one negative RGB channel. The reference column is the $N=24576$-direction numerical integration result.}
  \label{fig:vis}
\end{strip}

\subsection{Discussion}

On the DOF--accuracy trade-off, SH-2 (9 DOF) achieves the lowest error, five-parameter MaxEnt and stored QZH (5 DOF) are close, and curve-fit QZH and MaxEnt-4 (4 DOF) differ by scene. Positivity guarantees matter in practice: on high-sharpness spotlights QZH exhibits clear negatives while MaxEnt remains non-negative. MaxEnt may be suitable for local lighting representations with a single dominant direction and high concentration, but path guiding involves sampling PDFs, MIS, and spatiotemporal variation---its engineering value requires validation on real tasks. Regarding consistency with the original ZH3 paper, our current five-parameter subfamily fixes the dipole and second-order stress to be coaxial, covering only one of the axis strategies; the differences among independent-axis, shared-axis, and coaxial-projection strategies in error and color shift lie outside the scope of the present experiments.

\FloatBarrier   

\section{Limitations}

\subsection{Model Scope}

The closure presented in this paper is determined solely by moments up to $l\leq 2$. Although the full MaxEnt density generates constrained high-frequency bands, it cannot recover arbitrary unknown high-frequency structure present in the input; two real light fields with identical low-order moments will receive the same closure. The pure dipole model ignores second moments, and the coaxial five-parameter model requires the dipole and second-order stress to be coaxial; consequently, multi-peak, biaxial, or generally non-coaxial light fields may incur systematic closure error. The full 9-DOF model retains the general second moment, but its diffuse reconstruction and natural-parameter inversion are more expensive.

This work relies on a non-negative angular energy measure. Estimators with signed path weights, control variates, or negative-kernel filters cannot directly use the Jensen cone and probability density interpretation; one possible extension is to encode positive and negative measures separately, but its numerical stability and error propagation have not been studied.

\subsection{Numerical Implementation}

The mapping from natural parameters to moment parameters is unique in the interior of the realizable domain but becomes ill-conditioned near the boundary. The inner azimuthal integral has an algebraic closed form; the full diffuse response still involves an outer one-dimensional integral. The LUT moves this integral to an offline stage, at the cost of introducing grid, interpolation, and quantization errors.

The experiments chapter reports LUT interpolation error, half-precision quantization, boundary degenerate-region behavior, and GPU/CPU numerical differences; on the same 1024 evaluation normals as the main benchmark, the R16F $32^3$ LUT achieves a median benefit retention rate of 0.9997 with no observed negatives or NaN. The current implementation includes finite-precision safeguards; the triggering frequency of these guard branches has not been quantified separately.

\subsection{Experimental Evidence}

The 981 Poly Haven HDRIs cover outdoor, indoor, night, and sun scenes; the paired statistics and bucket analyses provide stable ranking evidence. Limitations of this dataset include: all are static environment lighting (no dynamics or temporal variation), the evaluation kernel is solely the Lambert cosine response (not covering BRDF-dependent error), and the coaxial assumption shared by the five-parameter model and QZH leads to systematic closure error on non-coaxial scenes such as tunnels and multi-source indoors. Both the current stored QZH and MaxEnt-5 use per-channel axes, lacking ablation on shared-axis and optimized-axis strategies.

\section{Conclusion}

Starting from the linear moments of energy-weighted directional samples, we construct a 9-scalar representation that preserves angular information up to $l\leq 2$, and under a fixed reference measure obtain a maximum-entropy closure with $g^{-3}$ and $g^{-4}$ as its core functional forms. The pure dipole subfamily provides analytic reconstruction and inverse sampling, and the coaxial five-parameter subfamily retains one independent second-order moment while enabling reconstruction via an offline LUT. The five-parameter state shares the same low-order axisymmetric moments as stored QZH, but replaces the finite-order polynomial truncation with a positive nonlinear closure.

On 981 Poly Haven HDRIs, five-parameter MaxEnt achieves a 78.7\% paired win rate against stored QZH, with the advantage increasing monotonically with lighting directionality, while maintaining zero negative values across all scenes; full second-order SH-2 remains superior on non-coaxial multi-source scenes. Future work includes shared-axis strategy ablation and full-GPU-load retesting.

\section*{Statement on AI-Assisted Tools}

Generative AI tools were used to assist with manuscript drafting, program implementation, and verification design. All mathematical arguments, experimental analyses, bibliographic references, and final wording were reviewed by the authors, who take full responsibility for this work.

\FloatBarrier   
\bibliographystyle{IEEEtran}
\bibliography{refs}

\end{document}